\pdfoutput=1

\documentclass[prb,amsmath,amssymb,amsbsy,twocolumn,superscriptaddress,showpacs]{revtex4-1}

\usepackage[svgnames]{xcolor}
\definecolor{darkblue}{RGB}{0 60 120}
\definecolor{eggplant}{RGB}{190 10 150}
\definecolor{darkgray}{RGB}{70 70 70}
\usepackage{amsfonts}
\usepackage{hhline}
\usepackage{amsmath}
\usepackage{amssymb}
\usepackage{bm}
\usepackage{graphicx}
\usepackage[breaklinks, colorlinks = true,linkcolor=eggplant,urlcolor=darkblue,citecolor=eggplant]{hyperref}
\usepackage{mathrsfs}
\usepackage[lofdepth,lotdepth,caption=false]{subfig}
\usepackage{varwidth}
\usepackage{wrapfig}
\usepackage{times}
\usepackage{bm}
\usepackage[percent]{overpic}
\usepackage{longtable}
\usepackage{multirow}
\newcommand{\liiro}{Li${}_{2}$IrO${}_{3}$}
\newcommand{\lio}{LIO}
\newcommand{\blio}{$\beta$-\lio{}}
\newcommand{\glio}{$\gamma$-\lio{}}

\newcommand{\jhalf}{$j_{\text{eff}}=1/2$}

\newcommand{\xybond}{$X/Y$}
\newcommand{\zbond}{$Z$}
\newcommand{\Bzbond}{$Z_2$}
\newcommand{\NBzbond}{$Z_1$}
\newcommand{\jkg}{$JK\Gamma$}
\newcommand{\avg}[1]{\langle #1 \rangle}

\begin{document}

\title{Two iridates, two models, two approaches:\\a comparative study
  on magnetism in 3D honeycomb materials}

\author{Eric Kin-Ho Lee}
\affiliation{Department of Physics and Center for Quantum Materials,
University of Toronto, Toronto, Ontario M5S 1A7, Canada.}
\author{Jeffrey G. Rau}
\affiliation{Department of Physics and Astronomy, University of
  Waterloo, Ontario, N2L 3G1, Canada}
\author{Yong Baek Kim}
\affiliation{Department of Physics and Center for Quantum Materials,
University of Toronto, Toronto, Ontario M5S 1A7, Canada.}
\affiliation{Canadian Institute for Advanced Research/Quantum Materials Program, Toronto, Ontario MSG 1Z8, Canada}
\affiliation{School of Physics, Korea Institute for Advanced Study, Seoul 130-722, Korea.}

\begin{abstract}
  Two recent theoretical works studied the role of Kitaev interactions
  in the newly observed incommensurate magnetic order in the
  hyper-honeycomb ($\beta$-Li$_2$IrO$_3$) and stripy-honeycomb
  ($\gamma$-Li$_2$IrO$_3$) iridates. Each of these works analyzed a
  different model (\jkg{} versus coupled zigzag chain model) using a
  contrasting method (classical versus soft-spin analysis).  The lack
  of commonality between these works precludes meaningful comparisons
  and a proper understanding of these unusual orderings.  In this
  study, we complete the unfinished picture initiated by these two
  works by solving \emph{both} models with \emph{both} approaches for
  \emph{both} 3D honeycomb iridates.  Through comparisons between all
  combinations of models, techniques, and materials, we find that the
  bond-isotropic \jkg{} model consistently predicts the experimental
  phase of $\beta$-Li$_2$IrO$_3$ regardless of the method used, while
  the experimental phase of $\gamma$-Li$_2$IrO$_3$ can be generated by
  the soft-spin approach with eigenmode mixing irrespective of the
  model used.  To gain further insights, we solve a 1D quantum
  spin-chain model related to both 3D models using the density matrix
  renormalization group method to form a benchmark.  We discover that
  in the 1D model, incommensurate correlations in the classical and
  soft-spin analysis survive in the quantum limit only in the presence
  of the symmetric-off-diagonal exchange $\Gamma$ found in the \jkg{}
  model.  The relevance of these results to the real materials are
  also discussed.
\end{abstract}
\date{\today}
\maketitle

\section{\label{sec:intro}Introduction}

\begin{table*}[htb!]
\begin{ruledtabular}
\begin{tabular}{cc|cccc}
  \multirow{2}{*}{Lattice}&\multirow{2}{*}{Experimental}&\multirow{2}{*}{Model}&\multirow{2}{*}{Classical}&\multicolumn{2}{c}{Soft-spin}\\
  & & & &no linear combination&with linear combination\\
  \hline
  \multirow{2}{*}{\blio{}}&\multirow{2}{*}{$\mathbf{(A_a,C_b,F_c)}$}&\jkg{}&$\mathbf{(A_a,C_b,F_c)}$&$\mathbf{(A_a,C_b,F_c)}$&$\mathbf{(A_a,C_b,F_c)}$\footnotemark[1]\\
  & &CZC&Commensurate&$(A_a,0,F_c)$&$\mathbf{(A_a,C_b,F_c)}$\\
  \hline
  \multirow{2}{*}{\glio{}}&\multirow{2}{*}{$\mathbf{(A_a,F_b,F_c)}$}&\jkg{}&$(A_a,C_b,F_c)$&$(A_a,C_b,F_c)$&$\mathbf{(A_a,F_b,F_c)}$\\
  & &CZC&Commensurate&$(A_a,0,F_c)$&$\mathbf{(A_a,F_b,F_c)}$\\
\end{tabular}
\end{ruledtabular}
\footnotetext[1]{Since the symmetry of the experimental spiral phase was already obtained without linear combination of multiple eigenmodes, no such mixing was performed.}
\caption{\label{tbl:summary}Summary of results for all combinations
  of model, lattice, and method.  The
  irreducible representations of the $a$, $b$, and $c$ components
  of the moments in the ground states are given in a shorthand notation.
  The entry is in bold if the irreducible representations correspond
  to the experimental phase, \emph{i.e.} the theoretical phase
  matches in symmetry with the observed ordering.
}
\end{table*}

Kitaev's honeycomb model, which hosts an exact $\mathbb{Z}_2$ spin
liquid\cite{kitaev2006anyons}, has been identified to play a crucial
role in the low-energy description of the quasi-two-dimensional
layered honeycomb iridates $A_2$IrO$_3$ ($A$=Li,
Na).\cite{jackeli2009mott, chaloupka2010kitaev, singh2012relevance} A
flurry of theoretical\cite{jiang2011possible, reuther2011finite,
  kimchi2011kitaev, you2012doping, schaffer2012quantum,
  price2013finite, rau2014generic, katukuri2014kitaev} and
experimental\cite{choi2012spin, ye2012direct, comin2012na,
  gretarsson2013magnetic, cao2013evolution, manni2014effect,
  knolle2014raman, chun2015direct} studies have since examined the
material's properties. Most recently, two three-dimensional (3D)
analogues of these layered materials have been
synthesized\cite{takayama2015hyperhoneycomb, modic2014realization},
giving hope to being the first realization of a 3D spin liquid.  These
discoveries have spurred many theoretical studies on the nature of
these spin-orbit coupled Mott insulators.\cite{mandal2009exactly,
  lee2014heisenberg, lee2014order, lee2014topological,
  nasu2014vaporization, lee2015theory, kimchi2014unified,
  kim2015predominance, schaffer2015topological} However, much like
their 2D analogues, these two
materials---$\beta$-Li$_2$IrO$_3$\cite{biffin2014unconventional} and
$\gamma$-Li$_2$IrO$_3$\cite{biffin2014noncoplanar} (hereafter \blio{}
and \glio{})---are magnetically ordered rather than exhibiting
spin-liquid behavior. To better understand the role of Kitaev physics
in these systems, their magnetic behavior must first be scrutinized.
As of late, two theoretical studies have attacked the problem from two
complementary directions:\cite{lee2015theory, kimchi2014unified} each
of these studies proposed a pseudospin 1/2 model describing the
low-energy physics of the two iridates and analyzed their respective
model using a different approach.  The first work examined the \jkg{}
model using a classical approach\cite{lee2015theory}, while the second
analyzed the coupled zigzag chain (CZC) model using a soft-spin
method\cite{kimchi2014unified} (see Sec. \ref{sec:models} for
definition of the models). The goal of both works is to utilize the
detailed experimental magnetic structure to constrain their respective
model to the region of the parameter space in which
$\beta$-Li$_2$IrO$_3$ and $\gamma$-Li$_2$IrO$_3$ reside.  One common
conclusion between the two studies is that the dominance of a
ferromagnetic Kitaev term and the presence of subdominant exchange
terms (including the antiferromagnetic Heisenberg exchange) generate
spiral orders through frustration arising from anisotropic exchanges.

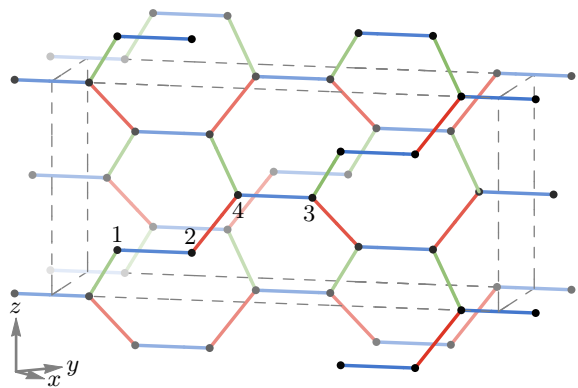
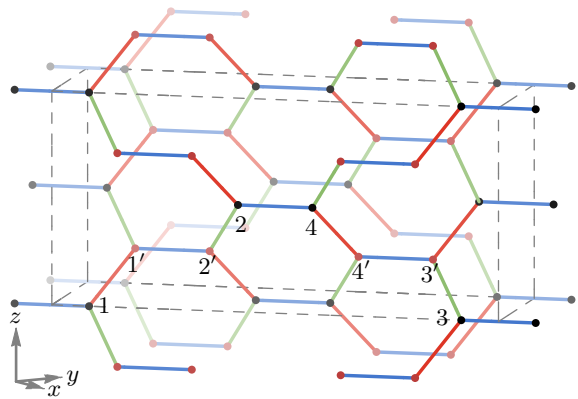
\begin{figure}[th!]
  \centering
  \setlength\fboxsep{0pt}
  \setlength\fboxrule{0pt}
  \subfloat[][Hyper-honeycomb lattice]{
    \label{fig:h0_lattice} 
    \fbox{\begin{overpic}[scale=1,clip=true,trim=0 45 0 45]{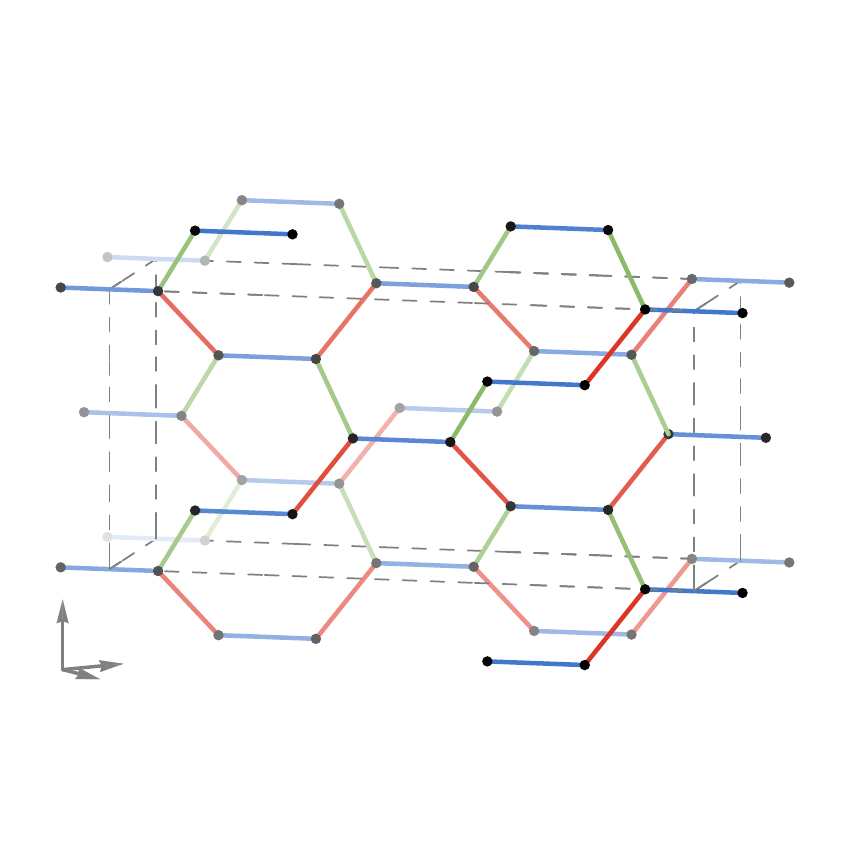}
        \put(12.2,.7){$\scriptsize{x}$}
        \put(15.1,3){$\scriptsize{y}$}
        \put(6.3,12){$\scriptsize{z}$}
        \put(22,22.9){$\scriptsize{1}$}
        \put(33.2,22.5){$\scriptsize{2}$}
        \put(40.6,26.6){$\scriptsize{4}$}
        \put(51.5,26.1){$\scriptsize{3}$}
      \end{overpic}}
  } 

  \subfloat[][Stripy-honeycomb lattice]{
    \label{fig:h1_lattice} 
    \fbox{\begin{overpic}[scale=1,clip=true,trim=0 45 0 45]{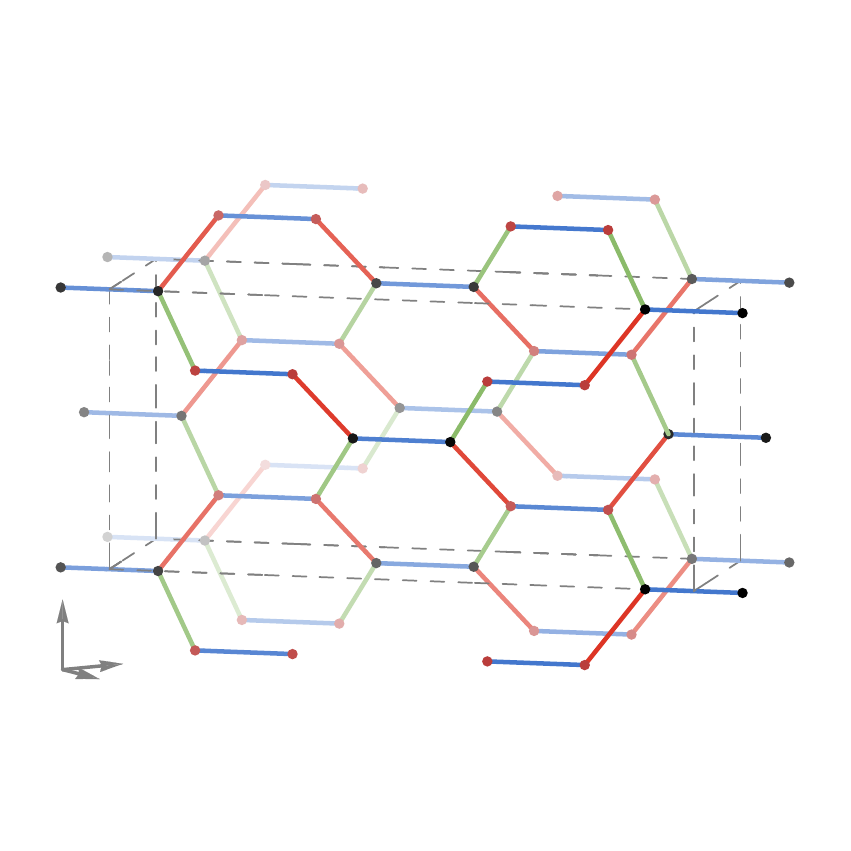}
        \put(12.2,.7){$\scriptsize{x}$}
        \put(15.1,3){$\scriptsize{y}$}
        \put(6.3,12){$\scriptsize{z}$}
        \put(24.4,19.8){$\scriptsize{1'}$}
        \put(35.4,19.5){$\scriptsize{2'}$}
        \put(59,18.7){$\scriptsize{4'}$}
        \put(69.6,18.2){$\scriptsize{3'}$}
        \put(20,13.5){$\scriptsize{1}$}
        \put(41,26){$\scriptsize{2}$}
        \put(51.8,25.5){$\scriptsize{4}$}
        \put(72.0,11.6){$\scriptsize{3}$}
    \end{overpic}}
  }
  \caption{\label{fig:lattices}(Color online) Lattices of the hyper-
    and stripy-honeycomb.  The red and green bonds are the
    symmetry-equivalent $X$ and $Y$ bonds.  The blue bonds are the $Z$
    bonds, which are symmetry-inequivalent to the \xybond{} bonds.
    For the stripy-honeycomb lattice, there are two inquivalent sets
    of $Z$ bonds.  The bonds between black sites are the \NBzbond{}
    bonds while the bonds between red sites are the \Bzbond{} bonds.
    The labeling of these bonds correspond to the Ising component
    present in the definition of the Kitaev interactions in both the
    \jkg{} and CZC models.  The numbers indicate the sublattice
    labeling used in this work.}
\end{figure}

While both works unearthed certain features of the magnetism in these
new iridates, their approaches and results have important differences.
Without a reliable method of solving a frustrated quantum spin model
in three dimensions, a definitive evaluation of these two models and
methods is beyond reach.  All hope is not lost, however: a careful
comparison of the models and techniques used can offer a much more
cohesive and encompassing understanding of the magnetism in these 3D
honeycomb iridates.

In this study, we complete the picture depicted in these previous
works by examining the \jkg{} \emph{and} CZC models using both
classical \emph{and} soft-spin approaches.  This work compares all
combinations of models and methods, culminating in an understanding
greater than the sum of its parts. Our main result is summarized in
Table \ref{tbl:summary}, where we compare the resulting ground states
of all eight combinations of materials, methods, and models.  With
these results, we conclude that the \jkg{} model on the
hyper-honeycomb is most robust as it predicts the experimental phase
under all methods.  Additionally, we find that the \glio{} magnetic
structure can only be captured with the soft-spin approach with linear
combination of multiple eigenmodes in both the \jkg{} and CZC models.

Ultimately, to understand how each of these methods succeeds or fails
to capture the quantum ground state properties, a comparison with a
quantum treatment would be highly desirable.  However, these
frustrated quantum mechanical spin models are difficult to solve in
three dimensions.  Hence, we analyze a 1D spin-chain model that is
related to both the \jkg{} and CZC models.  In one dimension, the
quantum solution of the spin model was obtained numerically using the
density matrix renormalization group (DMRG) technique.  We compare the
classical solution and soft-spin analysis with the quantum results and
find that incommensurate correlations found in both classical and
soft-spin methods only persist in the quantum limit when the symmetric
off-diagonal exchange $\Gamma$ is finite.  Although useful as an
illustrative tool, we also caution the extrapolation of these results
to higher dimensions, noting that quantum models can behave
drastically differently as the dimensionality of the system changes.

In Sec. \ref{sec:materials}, we briefly provide background information
on the two iridates in question: the hyper-honeycomb \blio{} and the
stripy-honeycomb \glio{}.  In Sec. \ref{sec:models}, we outline the
differences between the two spin models---the \jkg{} and the CZC
models.  Then, in Sec. \ref{sec:approaches}, we delve into the details
of the classical and soft-spin approaches, applying these methods to
obtain phase diagrams for the two models at hand.  We conclude this
section with a discussion about the implications of our findings.  We
present our 1D spin chain results in Sec. \ref{sec:1d}, and relate
these results to those obtained earlier for the 3D models.  Lastly in
Sec. \ref{sec:summary}, we provide a summary of our results and an
outlook on the topic of magnetism in these 3D iridate materials.

\section{\label{sec:materials}Two iridates}
The two 3D honeycomb iridates are structural variants (polymorphs) of
\liiro{}, and the elementary building blocks of both compounds is the
IrO$_6$ octahedron.  Each octahedron shares three edges with its three
neighbors, defining a tri-coordinated network of Ir ions.  The
resulting networks differ between the two compounds, as seen in
Fig. \ref{fig:lattices}.  Distinguishing features include the number
of symmetry-inequivalent nearest-neighbor (NN) bonds and the
symmetries they possess.  In the hyper-honeycomb, there are only two
symmetry-inequivalent sets of bonds: the \xybond{} bonds that
possesses inversion symmetry and the \zbond{} bonds that possesses
$222$ point group symmetry.  In contrast, there are three
symmetry-inequivalent sets of NN bonds in the stripy-honeycomb: the
\xybond{} bonds that do not possess any symmetry, the \NBzbond{} bonds
that possess $2/m$ symmetry, and the \Bzbond{} bonds that possess
$222$ symmetry. These distinctions are important in the construction
of models for these materials and in results discussed in
Sec. \ref{subsec:discussion}.\footnote{For further details on the
  crystal structure of these two 3D honeycomb iridates, we refer to
  earlier works that have elaborated on the description of the
  structure and crystal symmetries.\cite{modic2014realization,
    takayama2015hyperhoneycomb, lee2015theory}}

The magnetism found in these two iridates have many striking
similarities.  Both materials undergo a magnetic phase transition at
$T\sim 38~K$.\cite{takayama2015hyperhoneycomb, modic2014realization}
The magnetic order in both cases is a non-coplanar, incommensurate
spiral order, where moments on neighboring sublattices rotate in the
opposite sense (also known as \emph{counter-rotation} of spirals).
Moreover, the wavevector of both spirals are equal within experimental
resolution: $\sim$(0.57,0,0) in the $hkl$ notation of each compound's
respective orthorhombic unit cell.\cite{biffin2014unconventional,
  biffin2014noncoplanar}

While very similar, differences in the detailed magnetic structure have
been identified.\cite{biffin2014unconventional, biffin2014noncoplanar}
In the hyper-honeycomb case, the spiral order transforms under a
single irreducible representation (irrep) of the magnetic space group:
$\Gamma_4$ (see Appendix \ref{app:irrep} for
details).\cite{biffin2014unconventional} In other words, when applied
to the spiral order, all symmetries of the lattice that leaves the
ordering wavevector $\sim(0.57,0,0)$ invariant results in a factor of
$\pm1$, where the sign is dictated by the character of the irreducible
representation.  In contrast, the spiral in the stripy-honeycomb case
transforms in a more complicated fashion: the moments parallel to the
orthorhombic $b$ direction transform under a different irrep
($\Gamma_3$) than the moments perpendicular to the $b$ direction
($\Gamma_4$).\cite{biffin2014noncoplanar} Alternatively, in terms of
magnetic basis vectors, this implies that the magnetic structure of
\blio{} transforms as $(A_a,C_b,F_c)$ while the magnetic structure of
\glio{} transforms as $(A_a,F_b,F_c)$. This subtle difference between
the two spiral orders is outlined in the first two columns of Table
\ref{tbl:summary} and plays a crucial role in our discussion when
comparing the two models and the two methods.

\section{\label{sec:models}Two models}
A low-energy description of these Mott insulators with strong SOC can
be obtained by starting in the atomic picture.  In this limit, the
partially-filled Ir$^{4+}$ ions are in the $5d^5$ configuration.  In
the presence of large octahedral crystal fields and strong SOC, an
effective low-energy description can be derived from a single electron
occupying the high-energy \jhalf{} doublet.  In the Mott insulating
limit, the low-energy degrees of freedom of these half-filled doublets
can be represented by \jhalf{} pseudospins.  Due to the strong SOC,
these pseudospins are expected to interact via strongly anisotropic
exchanges (\emph{i.e.} strong exchange-anisotropy).  This is supported
by perturbative calculations that capture both the effects of virtual
states in the presence of Hund's coupling and oxygen-mediated
superexchange mechanisms.  We now explore two pseudospin models
motivated by these symmetry considerations.

\subsection{\label{subsec:jkg}\jkg{} model}
The \jkg{} model proposed in Ref. \onlinecite{lee2015theory} assumes
that all bonds have the same exchange parameters
(\textit{bond-isotropy}).  In addition, the model assumes all bonds
take on three exchange parameters $J$, $K$, and $\Gamma$, resulting in
the Hamiltonian
\begin{equation}
  \label{eq:ham_jkg}
  H=\sum_{\langle ij\rangle \in \alpha \beta (\gamma)}
  \left[J \vec{S}_i \cdot \vec{S}_j
    + K S^{\gamma}_{i} S^{\gamma}_{j}
    + e^{i\theta_{ij}}\Gamma S^{\alpha}_{i} S^{\beta}_{j}\right],
\end{equation}
where the pseudospin at site $i$ is denoted as $\vec{S}_i$, NN bonds
are denoted as $\langle ij \rangle$, and each NN bond is labelled by
$\gamma\in(X,Y,Z)$.  The shorthand $\langle ij \rangle \in \alpha
\beta (\gamma)$ means $\langle ij \rangle \in \gamma$ and an implicit
sum is taken over only $\alpha$ and $\beta$ where $\alpha \neq \beta
\neq \gamma$.  The phase angle $\theta_{ij}\in\{0,\pi\}\text{ mod }2\pi$
is bond-dependent and is partially constrained by the symmetry of the
lattice.  In the model considered in Ref. \onlinecite{lee2015theory}
and here, it is given by $\theta_{ij}=\pi(\hat{r}_{ij}\cdot
\vec{v}+1)$ where $\hat{r}_{ij}$ is the unit vector from site $i$ to
$j$,
$\vec{v}=\frac{1}{\sqrt{2}}(\frac{\vec{b}}{4}+\frac{\vec{c}}{6})$, and
$\vec{b}$, $\vec{c}$ are the conventional lattice vectors.

\subsection{\label{subsec:czc}Coupled zigzag chain model}
The coupled zigzag chain (CZC) model proposed in
Ref. \onlinecite{kimchi2014unified} assumes that the \xybond{} bonds
and \zbond{} are distinct (for \glio{}, \Bzbond{} and \NBzbond{} bonds
are assumed to possess the same exchange parameters).  Hence, this is
an intrinsically \textit{bond-anisotropic} model.  Furthermore, the
number of exchange parameters have been restricted: while the \zbond{}
bonds (\Bzbond{} and \NBzbond{} bonds in \glio{}) are assumed to have
three parameters similar to the \jkg{} model, the \xybond{} bonds
possess only two exchanges that parametrize the Heisenberg and Kitaev
exchanges.  These exchanges are related as the \xybond{} bonds and
\zbond{} bonds share the same $J$ and $K$ values via
\begin{equation*}
  \label{eq:ham_czc}
  H=\sum_{\langle ij \rangle \in \gamma} [ J\vec{S}_i \cdot \vec{S}_j + K S^{\gamma}_iS^{\gamma}_j] + \sum_{\langle ij \rangle \in Z}I_c(\hat{r}_{ij}\cdot \vec{S}_i)(\hat{r}_{ij} \cdot \vec{S}_j).
\end{equation*}
In the above, $\langle ij \rangle$ denotes NN sites, $J, K, I_c$ are
the Heisenberg, Kitaev, and Ising couplings respectively, $\gamma$ is
the Kitaev component and label on bond $ij$, $\hat{r}_{ij}$ is the
unit vector from site $i$ to $j$, and the Ising term only sums over
the \zbond{} bonds.  One can also re-write the exchanges in terms of
the $JK\Gamma$ parametrization on each bond to manifestly show the
bond-anisotropy
\begin{align*}
  &J_Z=J + \frac{1}{2}I_c && K_Z=K - \frac{1}{2}I_c  &&& \Gamma_Z=\frac{1}{2}I_c \\
  &J_{X/Y}=J && K_{X/Y}=K &&& \Gamma_{X/Y}=0,
\end{align*}
where the bond label of each exchange is indicated by the subscript.

\section{\label{sec:approaches}Two approaches}
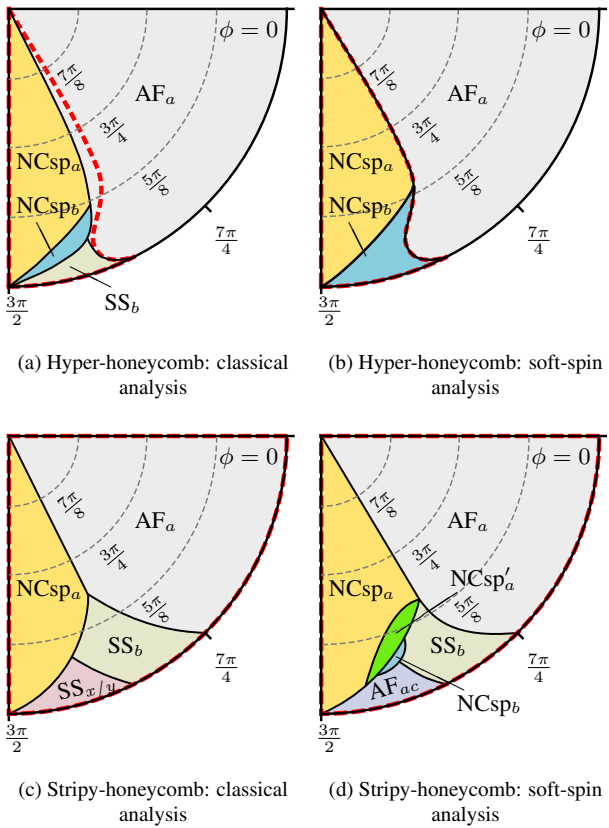
\begin{figure}[th!]
  \centering
  \setlength\fboxsep{0pt}
  \setlength\fboxrule{0pt}
  \subfloat[][Hyper-honeycomb: classical analysis]{
    \label{fig:h0_jkg_classical}
    \fbox{\begin{overpic}[scale=.93,clip=true,trim=0 0 -0 0]{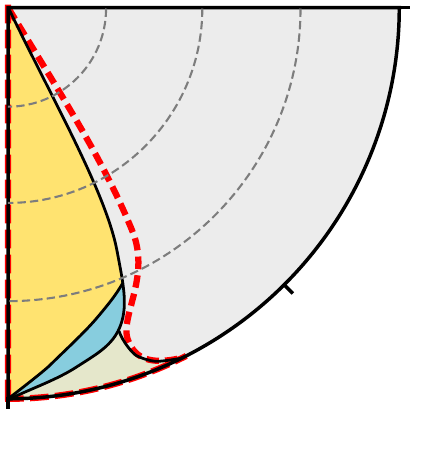}
        \put(64,90){$\scriptsize{\phi=0}$}
        \put(62,28){$\scriptsize{\frac{7 \pi}{4}}$}
        \put(0,7){$\scriptsize{\frac{3 \pi}{2}}$}
        \put(15,25){\color{black}\line(-1,2){5}}
        \put(20,22){\color{black}\line(2,-1){12}}
        \put(40,45){\rotatebox[origin=c]{40}{$\scriptsize{\frac{5 \pi}{8}}$}}
        \put(27.5,60.5){\rotatebox[origin=c]{40}{$\scriptsize{\frac{3 \pi}{4}}$}}
        \put(15,76){\rotatebox[origin=c]{40}{$\scriptsize{\frac{7 \pi}{8}}$}}
        \put(39,71){AF$_a$}
        \put(4,51){NCsp$_a$}
        \put(4,38){NCsp$_b$}
        \put(30,9){SS$_b$}
      \end{overpic}}
  }
  \subfloat[][Hyper-honeycomb: soft-spin analysis]{
    \label{fig:h0_jkg_soft}
    \fbox{\begin{overpic}[scale=.93,clip=true,trim=0 0 -0 0]{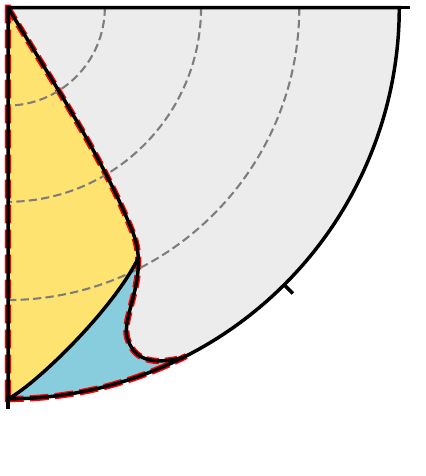}
        \put(64,90){$\scriptsize{\phi=0}$}
        \put(62,28){$\scriptsize{\frac{7 \pi}{4}}$}
        \put(0,7){$\scriptsize{\frac{3 \pi}{2}}$}
        \put(15,25){\color{black}\line(-1,2){5}}
        \put(40,45){\rotatebox[origin=c]{40}{$\scriptsize{\frac{5 \pi}{8}}$}}
        \put(27.5,60.5){\rotatebox[origin=c]{40}{$\scriptsize{\frac{3 \pi}{4}}$}}
        \put(15,76){\rotatebox[origin=c]{40}{$\scriptsize{\frac{7 \pi}{8}}$}}
        \put(39,71){AF$_a$}
        \put(4,51){NCsp$_a$}
        \put(4,38){NCsp$_b$}
      \end{overpic}}

  }
\\    
  \subfloat[][Stripy-honeycomb: classical analysis]{
    \label{fig:h1_jkg_classical}
    \fbox{\begin{overpic}[scale=.93,clip=true,trim=0 0 -0 0]{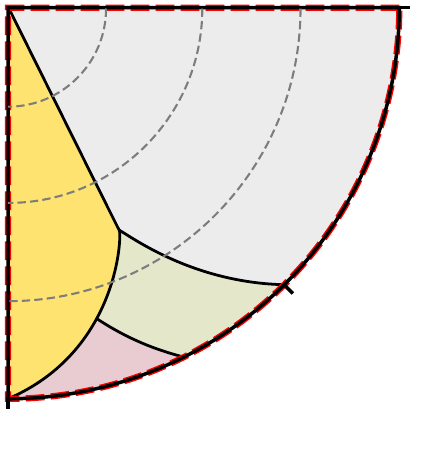}
        \put(64,90){$\scriptsize{\phi=0}$}
        \put(62,28){$\scriptsize{\frac{7 \pi}{4}}$}
        \put(0,7){$\scriptsize{\frac{3 \pi}{2}}$}
        \put(40,45){\rotatebox[origin=c]{40}{$\scriptsize{\frac{5 \pi}{8}}$}}
        \put(27.5,60.5){\rotatebox[origin=c]{40}{$\scriptsize{\frac{3 \pi}{4}}$}}
        \put(15,76){\rotatebox[origin=c]{40}{$\scriptsize{\frac{7 \pi}{8}}$}}
        \put(39,71){AF$_a$}
        \put(4,51){NCsp$_a$}
        \put(15.6,22.3){\rotatebox[origin=c]{15}{SS$_{x/y}$}}
        \put(31,34){SS$_b$}
      \end{overpic}}
  }
  \subfloat[][Stripy-honeycomb: soft-spin analysis]{
    \label{fig:h1_jkg_soft}
    \fbox{\begin{overpic}[scale=.93,clip=true,trim=0 0 -0 0]{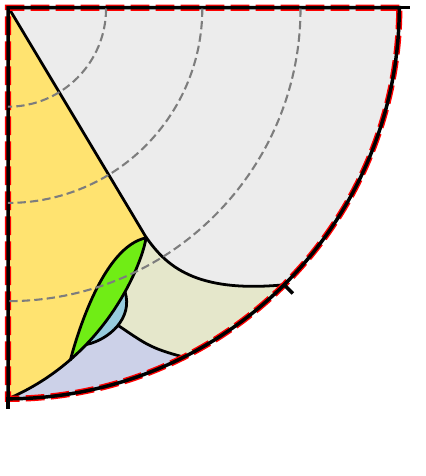}
        \put(64,90){$\scriptsize{\phi=0}$}
        \put(62,28){$\scriptsize{\frac{7 \pi}{4}}$}
        \put(0,7){$\scriptsize{\frac{3 \pi}{2}}$}
        \put(18,22){\color{black}\line(3,-1){7}}
        \put(24,40){\color{black}\line(3,2){18}}
        \put(24,33){\color{black}\line(2,-1){20}}
        \put(40,45){\rotatebox[origin=c]{40}{$\scriptsize{\frac{5 \pi}{8}}$}}
        \put(27.5,60.5){\rotatebox[origin=c]{40}{$\scriptsize{\frac{3 \pi}{4}}$}}
        \put(15,76){\rotatebox[origin=c]{40}{$\scriptsize{\frac{7 \pi}{8}}$}}
        \put(39,71){AF$_a$}
        \put(4,51){NCsp$_a$}
        \put(42,17){NCsp$_b$}
        \put(40,54){NCsp$_{a}'$}
        \put(15.2,22.1){\rotatebox[origin=c]{15}{AF$_{ac}$}}
        \put(34,34){SS$_b$}
      \end{overpic}}

  }
  \caption{\label{fig:jkg} (Color online) \jkg{} model: comparing the
    classical and soft-spin (with no mixing) approaches on the
    hyper-honeycomb and stripy-honeycomb lattices.  The classical
    approach stipulates the constant-spin-length constraint whereas
    the soft-spin approach does not.  LT fails in the region bounded
    by the red dotted line.  The non-coplanar spiral (NCsp$_a$) state
    exist in all phase diagrams---it is the experimentally observed
    phase in the hyper-honeycomb \blio{} and closely related to the
    experimental phase in the stripy-honeycomb \glio{}.  The values at
    the circle boundary are $\phi$-values, whereas values inside are
    $\theta$-values; see Eq. \ref{eq:parametrization} and surrounding
    main text for parametrization used.}
\end{figure}

\begin{figure}[th!]
  \centering
  \setlength\fboxsep{0pt}
  \setlength\fboxrule{0pt}
  \subfloat[][Hyper- and stripy-honeycomb: classical analysis]{
    \label{fig:h0_czc_classical} 
    \fbox{\begin{overpic}[scale=.93,clip=true,trim=0 0 0 0]{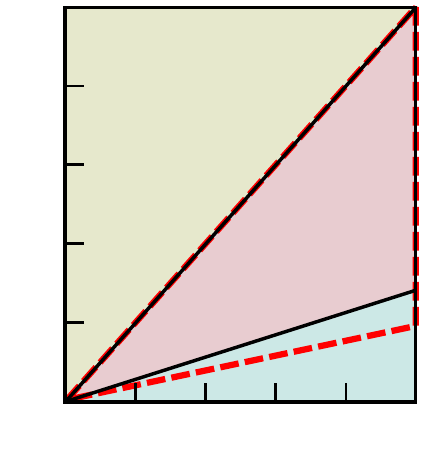}
        \put(.8,95){$\frac{J}{|K|}$}
        \put(84,8){$\frac{I_c}{K}$}
        \put(38,70){SS$_b$}
        \put(70,20){FM$_c$}
        \put(65,52){SS$_{x/y}$}
        \put(6,13.5){$0$}
        \put(1,30){$0.1$}
        \put(1,47){$0.2$}
        \put(1,63){$0.3$}
        \put(1,80){$0.4$}
        \put(12,8){$0$}
        \put(23,8){$0.2$}
        \put(38,8){$0.4$}
        \put(53,8){$0.6$}
        \put(68,8){$0.8$}
      \end{overpic}}
  } 
  \subfloat[][Hyper- and stripy-honeycomb: soft-spin analysis]{
    \label{fig:h0_czc_soft} 
    \fbox{\begin{overpic}[scale=.93,clip=true,trim=0 0 0 0]{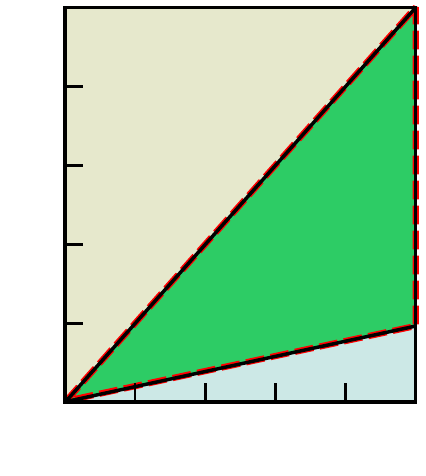}
        \put(.8,95){$\frac{J}{|K|}$}
        \put(84,8){$\frac{I_c}{K}$}
        \put(38,70){SS$_b$}
        \put(70,20){FM$_c$}
        \put(65,52){Csp$_{a}$}
        \put(6,13.5){$0$}
        \put(1,30){$0.1$}
        \put(1,47){$0.2$}
        \put(1,63){$0.3$}
        \put(1,80){$0.4$}
        \put(12,8){$0$}
        \put(23,8){$0.2$}
        \put(38,8){$0.4$}
        \put(53,8){$0.6$}
        \put(68,8){$0.8$}
    \end{overpic}}
  }
  \caption{\label{fig:h0_czc}(Color online) CZC model: comparing the
    classical and soft-spin (with no mixing) approaches on the
    hyper-honeycomb and stripy-honeycomb lattices.  The hyper- and
    stripy-honeycomb results agree within each method hence only one
    figure is shown per method.  LT fails in the region bounded by the
    red dotted line.  The classical method finds commensurate ordering
    in the whole parameter space considered, whereas the soft-spin
    approach finds a coplanar spiral in the LT-failed region.  See
    text for additional details on phases present and discussion on
    the comparison.}
\end{figure}
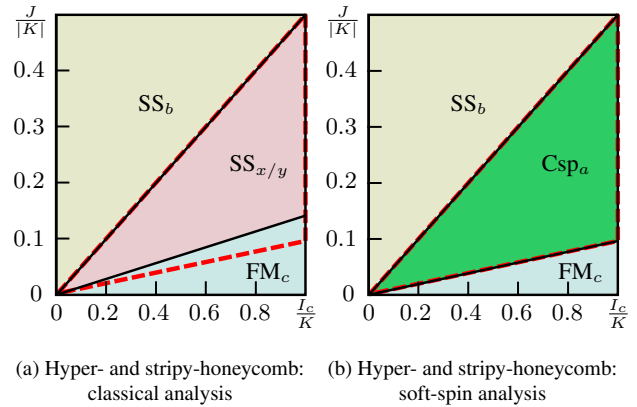

In the two studies Ref. \onlinecite{lee2015theory} and
\onlinecite{kimchi2014unified}, not only were the models considered
different, the approaches used to identify the ground states of each
model also differed.  Here we first provide details on both methods
then subsequently complete the analysis of the prior studies by
applying \emph{both} methods to \emph{both} models on the two iridates
in consideration.

\subsection{\label{subsec:classical}Classical analysis}
In the classical limit, the quantum mechanical pseudospins are treated
as constant length vectors.  For $S=1/2$, the classical treatment is
equivalent to a variational method where the ansatz is restricted to a
site-factorized product state.  This is because the product state
ansatz \emph{must} satisfy the classical unit length constraint due to
the unique Bloch sphere nature of $S=1/2$ spins: the expectation
value $\langle \vec{S} \rangle$ for any pure state forms a constant length
vector.  As such, to capture long-range ordered states, the classical
approach is a useful first method to employ.

To solve the classical problem (or equivalently, the variational
problem), we have various tools at our disposal.  The Luttinger-Tisza
(LT) approximation is an important method that serves to provide a
lower bound to the classical ground state energy and it also
identifies the exact classical ground state when the method succeeds.
Success is defined as finding a solution that satisfies the constant
spin-length constraint across all sites and is typically met in
regions of parameter space where geometrical and/or exchange
frustration is minimal.

Procedurally, the essence of the LT method involves taking a Fourier
transform of the real-space exchange Hamiltonian into momentum-space
and diagonalizing the resultant matrix at all momenta to yield an LT
band structure.  Then, we identify the momenta that globally minimize
such band structure and examine the corresponding degenerate
eigenspace.  Lastly, we attempt to construct an eigenvector within the
degenerate eigenspace that, when Fourier transformed back into
real-space, yields a constant spin-length configuration.  If such a
configuration is found, then the state is a classical ground state and
the energy corresponds to the state's eigenvalue.  If no such
configuration can be found, the LT method is deemed to have failed and
the lowest eigenvalue is the lower bound to the classical ground state
energy.  In the regions of parameter space where LT fails, we in turn
employ numerical methods such as simulated annealing to find the
classical ground state, supplemented with the knowledge of the lower
bound in the classical energy from the LT analysis.

\subsubsection{Previous results from \jkg{} model}
We first review the classical results of the \jkg{} model.  In
Fig. \ref{fig:h0_jkg_classical} and \ref{fig:h1_jkg_classical}, we
reproduce the classical phase diagram of the \jkg{} model for both 3D
honeycomb systems from Ref. \onlinecite{lee2015theory} with $J\ge 0$,
$K \le 0$, and $\Gamma \le 0$, which contains the experimentally
relevant spiral phases.  The exchange interactions are parametrized in
the polar plot as
\begin{equation}
  \label{eq:parametrization}
  (J,K,\Gamma)=(\sin \theta \cos \phi, \sin \theta \sin \phi, \cos \theta),
\end{equation}
where $\phi$ is the angular coordinate and $r=(\pi-\theta)$ is the
radial coordinate.  As such, the outer edge of the quarter-circle is
the Heisenberg-Kitaev limit, the left edge is the $J=0$ limit, and the
top edge is the $K=0$ limit.  Outside the red dotted lines, LT
succeeds and an antiferromagnetic phase (AF$_a$) exists.  Within the
red dotted lines, LT fails and we subsequently employed simulated
annealing to identify the ground states in this outlined region.

\paragraph*{Hyper-honeycomb:} The NCsp$_a$ (\emph{N}on-\emph{C}oplanar
\emph{sp}iral) phase in the dotted region reproduces the
experimentally observed magnetic order.  It is a non-coplanar,
counter-rotating spiral order with the experimentally observed broken
symmetries and an ordering wavevector in the $h00$ direction.  In
terms of the magnetic basis vectors, the NCsp$_a$ phase transforms as
$(A_a,C_b,F_c)$, \emph{i.e.} as a single-irrep $\Gamma_4$.  The
ordering wavevector in the NCsp$_a$ region continuously changes as a
function of $J$, $K$, and $\Gamma$, and the experimental wavevector is
contained within.  In addition to the NCsp$_a$ phase, there is also
another non-coplanar spiral with wavevector in the $0k0$ direction
(NCsp$_b$) and a skew-stripy phase (SS$_{b}$).

\paragraph*{Stripy-honeycomb:} In contrast to the hyper-honeycomb
case, none of the phases in the stripy-honeycomb phase diagram
reproduces the experimental phase exactly, although the NCsp$_a$ phase
in the stripy-honeycomb model is similar to the experimental phase.
In particular, the NCsp$_a$ phase transforms as the single irrep
$\Gamma_4$ with symmetry $(A_a,C_b,F_c)$, which differs from the
experimental phase in the $b$ component.  Like the hyper-honeycomb
case, the wavevectors in the NCsp$_a$ region continuously change as a
function of the exchange parameters and the region contains the
experimental wavevector.

For further details on the classical results for the \jkg{} model
including the existence of other magnetic orders and other parameter
regimes, we refer the reader to Ref. \onlinecite{lee2015theory}.

\subsubsection{New results from CZC model}
In Fig. \ref{fig:h0_czc_classical}, we applied the classical analysis
on the CZC model for both the hyper-honeycomb and stripy-honeycomb
lattices in the same parameter region as that of
Ref. \onlinecite{kimchi2014unified}, \emph{i.e.} $0\le J\le 0.5$,
$K=-1$, and $0\ge I_c\ge -1.0$.  The phase diagrams of both the hyper-
and stripy-honeycomb CZC model yield identical phase boundaries and
hence only one phase diagram is shown.  Outside the red dotted region
where either the Heisenberg exchange ($J$) or the Ising exchange
($I_c$) dominates, LT succeeds and identifies two distinct
commensurate phases.  In the large $J/|K|$ region, the ground state is
a skew-stripy phase (SS$_b$), while in the large $I_c/K$ region, a
collinear ferromagnetic ground state (FM$_c$) in the $c$ direction is
found.

In the dotted region where LT fails, simulated annealing identifies
commensurate ground states.  The FM$_c$ phase that was found in the
large $I_c/K$ region enlarges into the dotted region, and a new
skew-stripy phase (SS$_{a/b}$) emerges between the FM$_c$ and SS$_b$
phase.  The experimental spiral phase does not appear in the classical
analysis of the CZC model in either crystal structures.

\subsection{\label{subsec:soft-spin}Soft-spin analysis}

In the soft-spin analysis as proposed in
Ref. \onlinecite{kimchi2014unified}, one initially follows the same
procedure as the LT method as outlined in Sec. \ref{subsec:classical},
but disregards the constant spin-length constraint.  It was suggested
that the minimum eigenvalue solution, which may violate the
spin-length constraint, can be considered as a soft-spin
solution.\cite{kimchi2014unified} If the constant spin-length
condition is satisfied, however, the solution will be the same as the
classical solution.  It was further proposed that one can construct a
more complex spin structure with the same wavevector by forming a
linear combination, or \emph{mixing}, of the lowest and certain
higher-eigenvalue eigenvectors, which may be considered as a potential
candidate solution for the quantum model.\cite{kimchi2014unified} Here
we will follow their prescription and investigate the results in both
models.

\subsubsection{Previous results from CZC model}
In Fig. \ref{fig:h0_czc_soft}, we reproduced the soft-spin phase
diagram obtained in Ref. \onlinecite{kimchi2014unified}, which applies
to both hyper- and stripy-honeycomb CZC models.  As explained in
Sec. \ref{subsec:soft-spin}, the difference between the classical and
soft-spin results lies within the dotted region where LT fails.  In
this region, the soft-spin analysis finds the Csp$_a$ phase with a
continuously changing wavevector as a function of $J/|K|$ and $I_c/K$.
This phase is a coplanar, counter-rotating spiral with vanishing
moments along the orthorhombic $b$ direction and a wavevector along
the $h00$ direction.

This state differs from the experimental phase because moments are
only in the $a$-$c$ plane, hence yielding the symmetry $(A_a,0,F_c)$
($0$ is used to denote the vanishing $b$ component).  To obtain a
finite component along the $b$ direction while ensuring the symmetry
of the experimental phase, the eigenvector of the fifth-lowest
eigenvalue was linearly combined with the eigenvectors of the lowest
eigenvalue.\cite{kimchi2014unified} This mixed state has the symmetry
$(A_a,C_b,F_c)$ in the case of the hyper-honeycomb, and
$(A_a,F_b,F_c)$ in the case of the stripy-honeycomb, in agreement with
experimental results.\cite{kimchi2014unified} It was proposed that
this mixture of eigenvectors would result in a candidate solution for
the quantum model.\cite{kimchi2014unified}

\subsubsection{New results from \jkg{} model}
In Fig. \ref{fig:h0_jkg_soft} and \ref{fig:h1_jkg_soft}, we have
computed the phase diagram of the \jkg{} model using the soft-spin
approach. For both the hyper-honeycomb and stripy-honeycomb cases, in
the region where one would find the NCsp$_a$ phase using the classical
approach, we find phases with the same broken symmetry albeit with
modified ordering wavevector lengths and non-constant spin-lengths.
In this region of both lattices, the ordering wavevector continuously
changes and contains the experimentally verified wavevector.
 
\paragraph*{Hyper-honeycomb:} The symmetry of this lowest-eigenvalue
state is $(A_a,C_b,F_c)$, and since it agrees with the experimental
phase, a linear combination with higher-eigenvalue eigenvectors is not
necessary.  In other words, the \jkg{} model on the hyper-honeycomnb
lattice succeeds in finding the experimental phase in both the
classical analysis and the soft-spin analysis (without mixture).  In
comparison with the classical results, both the NCsp$_a$ and NCsp$_b$
phases enlarge and meets the boundary of where LT fails.  The SS$_b$
phase that was found in the classical analysis does not appear in the
soft-spin analysis.

\paragraph*{Stripy-honeycomb: } The lowest-eigenvalue state has
symmetry $(A_a,C_b,F_c)$, which is identical to the NCsp$_a$ phase in
the classical analysis.  A state with the experimental symmetry
$(A_a,F_b,F_c)$ can be obtained if higher-eigenvalue eigenvectors are
mixed, similar to the results outlined above for the CZC model.  In
addition to this phase, there are also two additional non-coplanar
spirals, NCsp$_{a}'$ and NCsp$_{b}$, and a non-coplanar, commensurate,
antiferromagnetic phase, AF$_{ac}$.  NCsp$_{a}'$ has an ordering
wavevector along the $h00$ direction but has a different symmetry from
NCsp$_a$.  On the other hand, NCsp$_b$ has an ordering wavevector
along the $0k0$ direction in analogy to the similarly named phase in
the hyper-honeycomb phase diagram.  The non-coplanar, commensurate
antiferromagnetic phase AF$_{ac}$ lies close to the Kitaev limit and
is distinct from the SS$_{x/y}$ phase predicted in the classical
analysis.  This phase consists of moments aligned
antiferromagnetically in the $c$ direction on the \NBzbond{} bonds,
and antiferromagnetically in the $a$ direction on the \Bzbond{} bonds
with a small ferromagnetic component along the $b$ axis.  Overall, the
phase does not have a net moment as the ferromagnetic tilts on the
various \Bzbond{} bonds within the unit cell are anti-aligned.

\subsection{\label{subsec:discussion}Discussion}
Table \ref{tbl:summary} summarizes our main results and allows for an
easy comparison between all methods, models, and lattices at a glance.
The table lists the symmetry of the experimentally observed ordering
for each lattice, followed by the symmetry of the generated phases
present for each model using classical analysis and the soft-spin
analysis with and without mixing of eigenmodes.

In the \blio{} case, the \jkg{} model generates the experimental phase
under all methods in consideration.  In contrast to the robustness of
the \jkg{} model, the CZC model can only generate the experimental
phase under the soft-spin approach with mixing of eigenmodes.  Since
the \jkg{} model is manifestly bond-isotropic while the CZC model is
not, this difference in robustness in generating the experimental
phase may be attributed to the bond-isotropy of the real material. The
bond-isotropy of \blio{} is supported by the near-ideal Ir-Ir and Ir-O
bond lengths and Ir-O-Ir bond angles\cite{takayama2015hyperhoneycomb}
in addition to the bond-isotropic orbital overlaps observed in recent
\emph{ab initio} results.\cite{kim2015predominance} For the \glio{}
case, the soft-spin approach with mixing of eigenmodes can generate
the experimental phase using both the \jkg{} and CZC models.  However,
none of the models can generate the experimental phase within the
classical approach or the soft-spin approach without linear
combination of eigenmodes.  This may also be related to the more
distorted nature of the \glio{} lattice as observed by structural
analysis,\cite{modic2014realization} which may imply large
bond-anisotropies in the effective spin model.  These anisotropies
together with the absence of any symmetry on the \xybond{} bonds
suggests that other symmetry-allowed, subdominant exchanges not yet
accounted for in either the \jkg{} or CZC model may be important in
determining the ground state in the real material.  To assess these
possibilities, \emph{ab initio} studies like DFT or quantum chemistry
calculations may provide the necessary insights and are thus crucial
in our understanding of \glio{}.

As noted in Sec. \ref{subsec:classical}, the classical solution, as
determined by LT or by any other method, yields a definite variational
wavefunction and variational energy of the quantum spin-$1/2$ model.
At the cost of obtaining a definite wavefunction and energy, the
classical approach ignores all quantum effects.  Although the
classical approach may correctly generate the broken symmetries deep
within the long-range ordered phases, near phase boundaries where
quantum fluctuations may be enhanced, this approach may fail to
identify the correct broken symmetries (if there are any) of the
ground state.  It may also be possible that the classical approach
fails completely when quantum fluctuations are sufficiently large that
the long-range order is destroyed.  Despite these shortcomings, some
quantum corrections to the wavefunction and energy in addition to the
stability of the phases can be partially accounted for via spin-wave
theory or Jastrow factors in variational Monte Carlo schemes.

It was suggested\cite{kimchi2014unified} that the soft-spin approach
may generate spin structures that incorporate the effects of quantum
fluctuations.  The procedure of forming linear combination of multiple
eigenmodes allows for construction of more complex ordering at the
same wavevector.  As a trade-off, the resulting expectation value of
the Hamiltonian using this soft-spin state (in both unmixed and mixed
cases) cannot be interpreted as a variatonal energy and the obtained
spin structure cannot be readily translated into a wavefunction.
Therefore, whether the classical or soft-spin result is favoured in a
quantum treatment at finite or zero temperature cannot be determined
within these two methods alone and a comparison with a quantum
treatment would be ideal.  However, the study of such 3D frustrated
quantum spin models is currently computationally prohibitive,
especially when searching for incommensurate phases.  Therefore, to
gather more insight regarding the methods and models considered thus
far, we consider a 1D spin chain limit of both the \jkg{} and CZC
models, which can be tackled by the DMRG technique.

\section{\label{sec:1d}1D spin chain model}

\begin{figure}[t!]
  \centering
  \setlength\fboxsep{0pt}
  \setlength\fboxrule{0pt}
  \subfloat[][DMRG - decoupled \jkg{}]{
    \label{fig:1d_jkg_dmrg} 
    \fbox{\begin{overpic}[scale=.93,clip=true,trim=0 0 0 0]{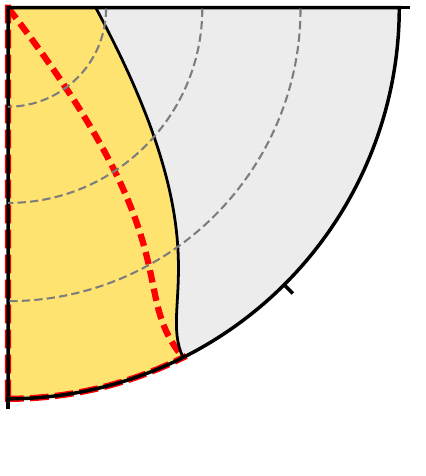}
        \put(64,90){$\scriptsize{\phi=0}$}
        \put(62,28){$\scriptsize{\frac{7 \pi}{4}}$}
        \put(0,7){$\scriptsize{\frac{3 \pi}{2}}$}
        \put(40,45){\rotatebox[origin=c]{40}{$\scriptsize{\frac{5 \pi}{8}}$}}
        \put(27.5,60.5){\rotatebox[origin=c]{40}{$\scriptsize{\frac{3 \pi}{4}}$}}
        \put(15,76){\rotatebox[origin=c]{40}{$\scriptsize{\frac{7 \pi}{8}}$}}
        \put(12,50){Sp}
        \put(45,73){AF}
      \end{overpic}}
  } 
  \subfloat[][DMRG - decoupled CZC]{
    \label{fig:1d_jkg_lassical} 
    \fbox{\begin{overpic}[scale=.93,clip=true,trim=0 0 0 0]{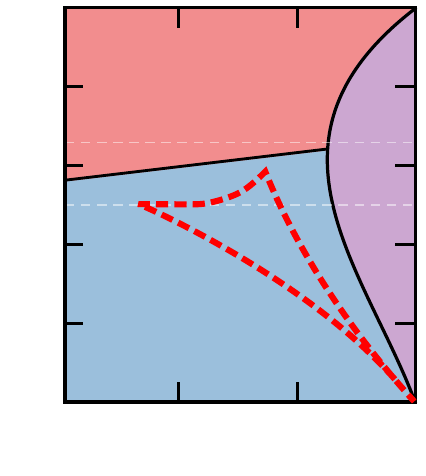}
        \put(.8,93){$\frac{J_1}{|K|}$}
        \put(80,8){$\frac{J_2}{|K|}$}
        \put(49,71){a}
        \put(49,51){b}
        \put(46,79){AF}
        \put(46,30){FM}
        \put(72,63){ST}
        \put(6,15){$0$}
        \put(1,30){$0.1$}
        \put(1,47){$0.2$}
        \put(1,63){$0.3$}
        \put(1,80){$0.4$}
        \put(3,8){$-0.24$}
        \put(28,8){$-0.16$}
        \put(53,8){$-0.08$}
      \end{overpic}}
  } 

  \subfloat[][Classical - decoupled \jkg{}]{
    \label{fig:1d_jkg_lassical} 
    \fbox{\begin{overpic}[scale=.93,clip=true,trim=0 0 0 0]{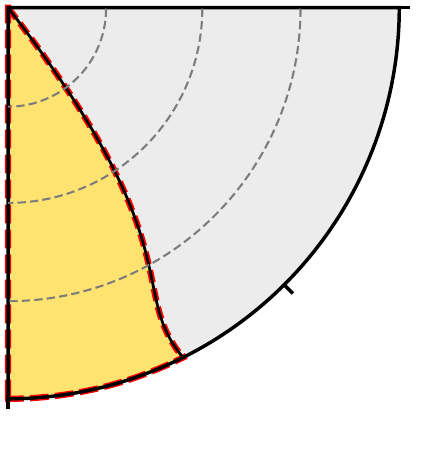}
        \put(64,90){$\scriptsize{\phi=0}$}
        \put(62,28){$\scriptsize{\frac{7 \pi}{4}}$}
        \put(0,7){$\scriptsize{\frac{3 \pi}{2}}$}
        \put(40,45){\rotatebox[origin=c]{40}{$\scriptsize{\frac{5 \pi}{8}}$}}
        \put(27.5,60.5){\rotatebox[origin=c]{40}{$\scriptsize{\frac{3 \pi}{4}}$}}
        \put(15,76){\rotatebox[origin=c]{40}{$\scriptsize{\frac{7 \pi}{8}}$}}
        \put(12,50){Sp}
        \put(45,73){AF}
    \end{overpic}}
  }
  \subfloat[][Classical - decoupled CZC]{
    \label{fig:1d_czc_classical} 
    \fbox{\begin{overpic}[scale=.93,clip=true,trim=0 0 0 0]{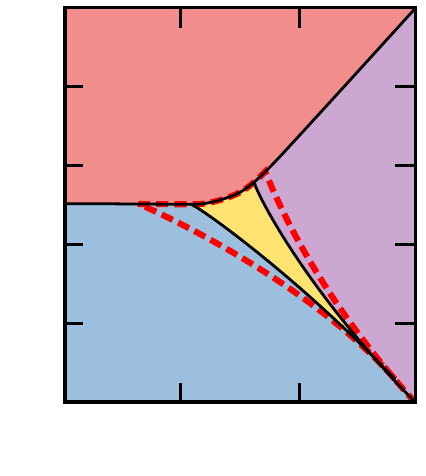}
        \put(.8,93){$\frac{J_1}{|K|}$}
        \put(80,8){$\frac{J_2}{|K|}$}
        \put(36,74){AF}
        \put(36,30){FM}
        \put(66,56){ST}
        \put(48,52){Sp}
        \put(6,15){$0$}
        \put(1,30){$0.1$}
        \put(1,47){$0.2$}
        \put(1,63){$0.3$}
        \put(1,80){$0.4$}
        \put(3,8){$-0.24$}
        \put(28,8){$-0.16$}
        \put(53,8){$-0.08$}
    \end{overpic}}
  }

  \subfloat[][Soft-spin - decoupled \jkg{}]{
    \label{fig:1d_jkg_soft} 
    \fbox{\begin{overpic}[scale=.93,clip=true,trim=0 0 0 0]{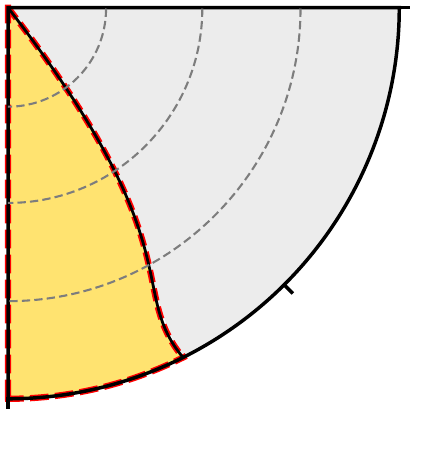}
        \put(64,90){$\scriptsize{\phi=0}$}
        \put(62,28){$\scriptsize{\frac{7 \pi}{4}}$}
        \put(0,7){$\scriptsize{\frac{3 \pi}{2}}$}
        \put(40,45){\rotatebox[origin=c]{40}{$\scriptsize{\frac{5 \pi}{8}}$}}
        \put(27.5,60.5){\rotatebox[origin=c]{40}{$\scriptsize{\frac{3 \pi}{4}}$}}
        \put(15,76){\rotatebox[origin=c]{40}{$\scriptsize{\frac{7 \pi}{8}}$}}
        \put(12,50){Sp}
        \put(45,73){AF}
    \end{overpic}}
  }
  \subfloat[][Soft-spin - decoupled CZC]{
    \label{fig:1d_czc_soft} 
    \fbox{\begin{overpic}[scale=.93,clip=true,trim=0 0 0 0]{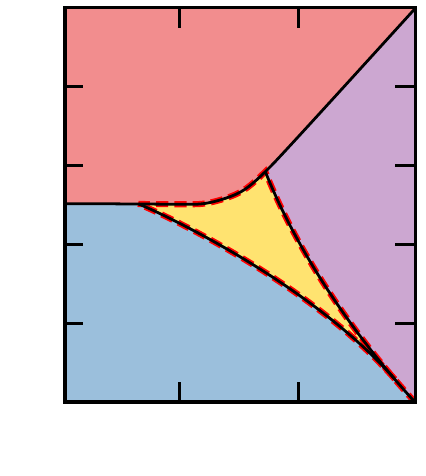}
        \put(.8,93){$\frac{J_1}{|K|}$}
        \put(80,8){$\frac{J_2}{|K|}$}
        \put(36,74){AF}
        \put(36,30){FM}
        \put(66,56){ST}
        \put(48,52){Sp}
        \put(6,15){$0$}
        \put(1,30){$0.1$}
        \put(1,47){$0.2$}
        \put(1,63){$0.3$}
        \put(1,80){$0.4$}
        \put(3,8){$-0.24$}
        \put(28,8){$-0.16$}
        \put(53,8){$-0.08$}
    \end{overpic}}
  }

  \caption{\label{fig:1d_jkg}(Color online) Phase diagram of 1D
    spin-chain model in the decoupled \jkg{} limit ($\Gamma<0$,
    $J_2=0$) and decoupled CZC limit ($\Gamma=0$, $J_2<0$) obtained
    via three different methods.  The regions enclosed by the red
    dotted lines indicate that LT failed. In a), c), and e), we
    studied the decoupled \jkg{} limit with DMRG, classical, and
    soft-spin analysis respectively.  Similarly, in b), d), and f), we
    simulated the decoupled CZC limit.  In the decoupled \jkg{} limit,
    all three methods predict an incommensurate phase (Sp) and an
    antiferromagnetic (AF) phase.  These phases are found in analogous
    positions in parameter space compared to the 3D \jkg{} models in
    Fig. \ref{fig:jkg}.  In the decoupled CZC limit, there are three
    commensurate phases: away from the stripy (ST) phase at small
    $J_2$, a ferromagnet (FM) and antiferromagnet (AF) appear.  The
    classical d) and soft-spin f) analyses also predict an
    intermediate incommensurate spiral (Sp) phase that does not appear
    in the quantum analysis.  The dotted lines in b) are cuts for
    which we show the structure factor in Fig. \ref{fig:dmrg_corr};
    the structure factor peaks for $\theta=\frac{3\pi}{4}$ in a) are
    shown in Fig. \ref{fig:1d_jkg_cut}.}
\end{figure}
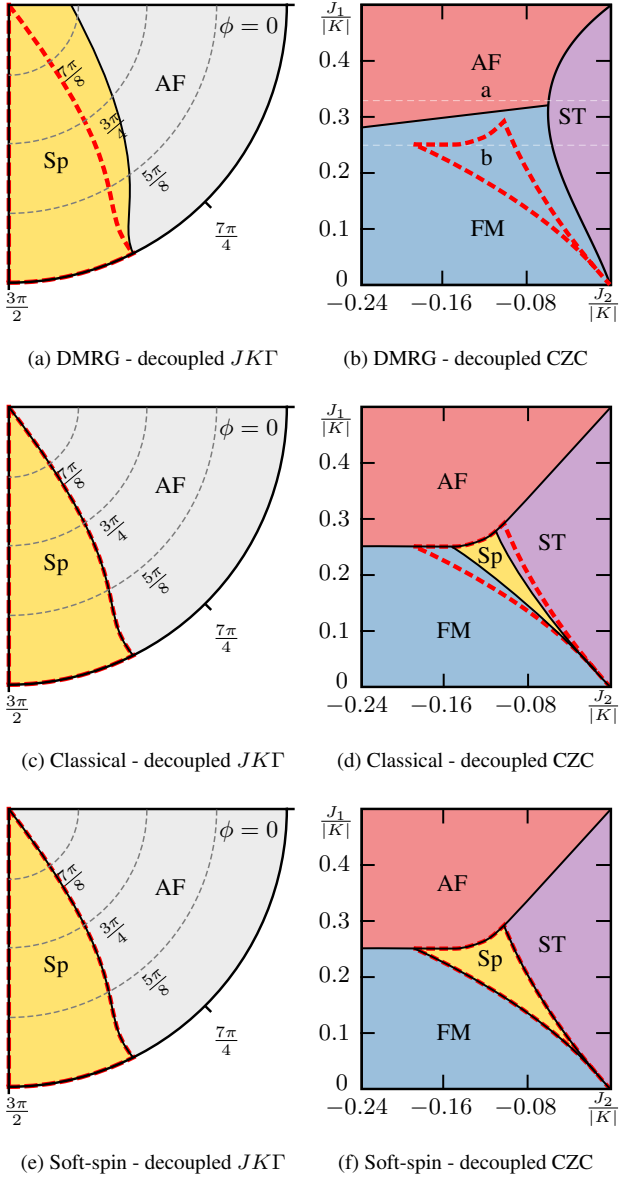

The 1D spin chain model related to both the \jkg{} and CZC models is
an extension of the spin chain model considered in
Ref. \onlinecite{kimchi2014unified}.  It is given by
\begin{align}
  \label{eq:chain}
  H=&\sum_{\langle ij\rangle \in \alpha \beta (\gamma)}
  J_1 \vec{S}_i \cdot \vec{S}_j
    + K S^{\gamma}_{i} S^{\gamma}_{j}
    + e^{i\theta_{ij}}\Gamma S^{\alpha}_{i} S^{\beta}_{j} \nonumber \\
  &+\sum_{\avg{\avg{ij}}}J_2  \vec{S}_i \cdot \vec{S}_j,
\end{align}
where only $X$ and $Y$ bonds are included in the sums,
$\avg{\avg{ij}}$ denotes second nearest neighbors (2NN), and the phase
factor appearing before $\Gamma$ is defined identically to
Eq. \ref{eq:ham_jkg}.  When $\Gamma\neq 0$ and $J_2=0$, we arrive at
the decoupled limit of the \jkg{} model: it is identical to
Eq. \ref{eq:ham_jkg} with vanishing $Z$ bond exchanges, resulting in
decoupled chains with only \xybond{} bonds.  When $\Gamma=0$ and
$J_2=0$, we similarly reach the decoupled limt of the CZC model.  The
2NN Heisenberg exchange has been introduced to this decoupled CZC
limit in Ref. \onlinecite{kimchi2014unified} to compensate for the
loss of the $Z$ bond exchanges.\cite{kimchi2014unified} By studying
these two limits with all three approaches---DMRG, classical, and
soft-spin analysis---we aim to compare the three methods and to
understand the essential ingredient needed to generate incommensurate
correlations.  A more in-depth discussion of the DMRG results can be
found in Appendix \ref{app:1d}.

In the decoupled \jkg{} limit, Fig. \ref{fig:1d_jkg} shows that all
three approaches capture incommensurate correlations.  This shows that
the spiral correlations generated by the frustration between the
$\Gamma$ term and the HK terms persist in the quantum limit.  We
notice that the classical and soft-spin analysis yield similar phase
boundaries that coincide with the area where LT fails (enclosed by the
red dotted lines), while in the quantum limit, the incommensurate
region enlarges beyond the red dotted lines.  On the contrary, in the
decoupled CZC limit, we find that the quantum phase diagram obtained
via DMRG does not contain the incommensurate correlations predicted by
both the classical and soft-spin analysis.  In particular, the
commensurate phases are captured by all three methods (quantum,
classical, and soft-spin), but an intermediate spiral phase is only
present in the classical and soft-spin analysis within the region
where LT fails.  This suggests that the incommensurate correlations
driven by the 2NN Heisenberg is ultimately disfavoured by quantum
effects.  Taken together, these two sets of results suggests that
finite $\Gamma$, and not further neighbor interactions, can stablize
incommensurate correlations in the 1D limit.  

One may hypothesize that the same trend will occur in the 3D models
discussed in this work: in the quantum limit, the Csp$_a$ phase
predicted using the soft-spin approach in the CZC model for both
lattices may be disfavoured relative to the commensurate phases due to
the absence of $\Gamma$ along the chain direction.  However, since
quantum effects in 1D systems are typically greater than 3D systems,
this extrapolation of results remains speculative.  Future studies are
needed to establish the connection between these 1D decoupled chain
results and the quantum phase diagram of the 3D models.

\section{\label{sec:summary}Summary}
We have completed the picture initated in two previous works by
examining both the \jkg{} and CZC models using both the classical and
soft-spin approaches.  For the hyper-honeycomb, the robustness of the
\jkg{} model in predicting the experimental phase regardless of the
method used suggests that bond-isotropic interactions can explain the
observed spiral order.  On the other hand, the experimental phase of
\glio{} can only be reproduced using the soft-spin analysis with
mixing of eigenmodes using both \jkg{} and CZC models, suggesting that
further investigations are needed to deduce an effective model from
microscopic origins.  In all cases where the experimental phase was
predicted, a dominant ferromagnetic Kitaev exchange exists as a common
thread.  We conclude by examining the decoupled limit of both the
\jkg{} and CZC models and observed that a finite $\Gamma$ was needed
to ensure that the incommensurate correlations predicted in both the
classical and soft-spin methods persist in the quantum limit.  Further
exploration extending these findings to higher dimensions are
warranted in order to assess the relevance of these results to the 3D
honeycomb iridates.

\acknowledgements We thank I. Kimchi, A. Vishwanath, and R. Coldea for
helpful discussions.  We are very grateful to R. Coldea for explaining
his experimental data in great detail. Computations were performed on
the GPC supercomputer at the SciNet HPC Consortium. SciNet is funded
by: the Canada Foundation for Innovation under the auspices of Compute
Canada; the Government of Ontario; Ontario Research Fund - Research
Excellence; and the University of Toronto.  This research was
supported by the NSERC, CIFAR, and Centre for Quantum Materials at the
University of Toronto.

\bibliography{comp_study}

\begin{thebibliography}{35}%
\makeatletter
\providecommand \@ifxundefined [1]{%
 \@ifx{#1\undefined}
}%
\providecommand \@ifnum [1]{%
 \ifnum #1\expandafter \@firstoftwo
 \else \expandafter \@secondoftwo
 \fi
}%
\providecommand \@ifx [1]{%
 \ifx #1\expandafter \@firstoftwo
 \else \expandafter \@secondoftwo
 \fi
}%
\providecommand \natexlab [1]{#1}%
\providecommand \enquote  [1]{``#1''}%
\providecommand \bibnamefont  [1]{#1}%
\providecommand \bibfnamefont [1]{#1}%
\providecommand \citenamefont [1]{#1}%
\providecommand \href@noop [0]{\@secondoftwo}%
\providecommand \href [0]{\begingroup \@sanitize@url \@href}%
\providecommand \@href[1]{\@@startlink{#1}\@@href}%
\providecommand \@@href[1]{\endgroup#1\@@endlink}%
\providecommand \@sanitize@url [0]{\catcode `\\12\catcode `\$12\catcode
  `\&12\catcode `\#12\catcode `\^12\catcode `\_12\catcode `\%12\relax}%
\providecommand \@@startlink[1]{}%
\providecommand \@@endlink[0]{}%
\providecommand \url  [0]{\begingroup\@sanitize@url \@url }%
\providecommand \@url [1]{\endgroup\@href {#1}{\urlprefix }}%
\providecommand \urlprefix  [0]{URL }%
\providecommand \Eprint [0]{\href }%
\providecommand \doibase [0]{http://dx.doi.org/}%
\providecommand \selectlanguage [0]{\@gobble}%
\providecommand \bibinfo  [0]{\@secondoftwo}%
\providecommand \bibfield  [0]{\@secondoftwo}%
\providecommand \translation [1]{[#1]}%
\providecommand \BibitemOpen [0]{}%
\providecommand \bibitemStop [0]{}%
\providecommand \bibitemNoStop [0]{.\EOS\space}%
\providecommand \EOS [0]{\spacefactor3000\relax}%
\providecommand \BibitemShut  [1]{\csname bibitem#1\endcsname}%
\let\auto@bib@innerbib\@empty
\bibitem [{\citenamefont {Kitaev}(2006)}]{kitaev2006anyons}%
  \BibitemOpen
  \bibfield  {author} {\bibinfo {author} {\bibfnamefont {A.}~\bibnamefont
  {Kitaev}},\ }\href {\doibase http://dx.doi.org/10.1016/j.aop.2005.10.005}
  {\bibfield  {journal} {\bibinfo  {journal} {Annals of Physics}\ }\textbf
  {\bibinfo {volume} {321}},\ \bibinfo {pages} {2 } (\bibinfo {year}
  {2006})}\BibitemShut {NoStop}%
\bibitem [{\citenamefont {Jackeli}\ and\ \citenamefont
  {Khaliullin}(2009)}]{jackeli2009mott}%
  \BibitemOpen
  \bibfield  {author} {\bibinfo {author} {\bibfnamefont {G.}~\bibnamefont
  {Jackeli}}\ and\ \bibinfo {author} {\bibfnamefont {G.}~\bibnamefont
  {Khaliullin}},\ }\href {\doibase 10.1103/PhysRevLett.102.017205} {\bibfield
  {journal} {\bibinfo  {journal} {Phys. Rev. Lett.}\ }\textbf {\bibinfo
  {volume} {102}},\ \bibinfo {pages} {017205} (\bibinfo {year}
  {2009})}\BibitemShut {NoStop}%
\bibitem [{\citenamefont {Chaloupka}\ \emph {et~al.}(2010)\citenamefont
  {Chaloupka}, \citenamefont {Jackeli},\ and\ \citenamefont
  {Khaliullin}}]{chaloupka2010kitaev}%
  \BibitemOpen
  \bibfield  {author} {\bibinfo {author} {\bibfnamefont {J.~c.~v.}\
  \bibnamefont {Chaloupka}}, \bibinfo {author} {\bibfnamefont {G.}~\bibnamefont
  {Jackeli}}, \ and\ \bibinfo {author} {\bibfnamefont {G.}~\bibnamefont
  {Khaliullin}},\ }\href {\doibase 10.1103/PhysRevLett.105.027204} {\bibfield
  {journal} {\bibinfo  {journal} {Phys. Rev. Lett.}\ }\textbf {\bibinfo
  {volume} {105}},\ \bibinfo {pages} {027204} (\bibinfo {year}
  {2010})}\BibitemShut {NoStop}%
\bibitem [{\citenamefont {Singh}\ \emph {et~al.}(2012)\citenamefont {Singh},
  \citenamefont {Manni}, \citenamefont {Reuther}, \citenamefont {Berlijn},
  \citenamefont {Thomale}, \citenamefont {Ku}, \citenamefont {Trebst},\ and\
  \citenamefont {Gegenwart}}]{singh2012relevance}%
  \BibitemOpen
  \bibfield  {author} {\bibinfo {author} {\bibfnamefont {Y.}~\bibnamefont
  {Singh}}, \bibinfo {author} {\bibfnamefont {S.}~\bibnamefont {Manni}},
  \bibinfo {author} {\bibfnamefont {J.}~\bibnamefont {Reuther}}, \bibinfo
  {author} {\bibfnamefont {T.}~\bibnamefont {Berlijn}}, \bibinfo {author}
  {\bibfnamefont {R.}~\bibnamefont {Thomale}}, \bibinfo {author} {\bibfnamefont
  {W.}~\bibnamefont {Ku}}, \bibinfo {author} {\bibfnamefont {S.}~\bibnamefont
  {Trebst}}, \ and\ \bibinfo {author} {\bibfnamefont {P.}~\bibnamefont
  {Gegenwart}},\ }\href {\doibase 10.1103/PhysRevLett.108.127203} {\bibfield
  {journal} {\bibinfo  {journal} {Phys. Rev. Lett.}\ }\textbf {\bibinfo
  {volume} {108}},\ \bibinfo {pages} {127203} (\bibinfo {year}
  {2012})}\BibitemShut {NoStop}%
\bibitem [{\citenamefont {Jiang}\ \emph {et~al.}(2011)\citenamefont {Jiang},
  \citenamefont {Gu}, \citenamefont {Qi},\ and\ \citenamefont
  {Trebst}}]{jiang2011possible}%
  \BibitemOpen
  \bibfield  {author} {\bibinfo {author} {\bibfnamefont {H.-C.}\ \bibnamefont
  {Jiang}}, \bibinfo {author} {\bibfnamefont {Z.-C.}\ \bibnamefont {Gu}},
  \bibinfo {author} {\bibfnamefont {X.-L.}\ \bibnamefont {Qi}}, \ and\ \bibinfo
  {author} {\bibfnamefont {S.}~\bibnamefont {Trebst}},\ }\href@noop {}
  {\bibfield  {journal} {\bibinfo  {journal} {Physical Review B}\ }\textbf
  {\bibinfo {volume} {83}},\ \bibinfo {pages} {245104} (\bibinfo {year}
  {2011})}\BibitemShut {NoStop}%
\bibitem [{\citenamefont {Reuther}\ \emph {et~al.}(2011)\citenamefont
  {Reuther}, \citenamefont {Thomale},\ and\ \citenamefont
  {Trebst}}]{reuther2011finite}%
  \BibitemOpen
  \bibfield  {author} {\bibinfo {author} {\bibfnamefont {J.}~\bibnamefont
  {Reuther}}, \bibinfo {author} {\bibfnamefont {R.}~\bibnamefont {Thomale}}, \
  and\ \bibinfo {author} {\bibfnamefont {S.}~\bibnamefont {Trebst}},\ }\href
  {\doibase 10.1103/PhysRevB.84.100406} {\bibfield  {journal} {\bibinfo
  {journal} {Phys. Rev. B}\ }\textbf {\bibinfo {volume} {84}},\ \bibinfo
  {pages} {100406} (\bibinfo {year} {2011})}\BibitemShut {NoStop}%
\bibitem [{\citenamefont {Kimchi}\ and\ \citenamefont
  {You}(2011)}]{kimchi2011kitaev}%
  \BibitemOpen
  \bibfield  {author} {\bibinfo {author} {\bibfnamefont {I.}~\bibnamefont
  {Kimchi}}\ and\ \bibinfo {author} {\bibfnamefont {Y.-Z.}\ \bibnamefont
  {You}},\ }\href {\doibase 10.1103/PhysRevB.84.180407} {\bibfield  {journal}
  {\bibinfo  {journal} {Phys. Rev. B}\ }\textbf {\bibinfo {volume} {84}},\
  \bibinfo {pages} {180407} (\bibinfo {year} {2011})}\BibitemShut {NoStop}%
\bibitem [{\citenamefont {You}\ \emph {et~al.}(2012)\citenamefont {You},
  \citenamefont {Kimchi},\ and\ \citenamefont {Vishwanath}}]{you2012doping}%
  \BibitemOpen
  \bibfield  {author} {\bibinfo {author} {\bibfnamefont {Y.-Z.}\ \bibnamefont
  {You}}, \bibinfo {author} {\bibfnamefont {I.}~\bibnamefont {Kimchi}}, \ and\
  \bibinfo {author} {\bibfnamefont {A.}~\bibnamefont {Vishwanath}},\ }\href
  {\doibase 10.1103/PhysRevB.86.085145} {\bibfield  {journal} {\bibinfo
  {journal} {Phys. Rev. B}\ }\textbf {\bibinfo {volume} {86}},\ \bibinfo
  {pages} {085145} (\bibinfo {year} {2012})}\BibitemShut {NoStop}%
\bibitem [{\citenamefont {Schaffer}\ \emph {et~al.}(2012)\citenamefont
  {Schaffer}, \citenamefont {Bhattacharjee},\ and\ \citenamefont
  {Kim}}]{schaffer2012quantum}%
  \BibitemOpen
  \bibfield  {author} {\bibinfo {author} {\bibfnamefont {R.}~\bibnamefont
  {Schaffer}}, \bibinfo {author} {\bibfnamefont {S.}~\bibnamefont
  {Bhattacharjee}}, \ and\ \bibinfo {author} {\bibfnamefont {Y.~B.}\
  \bibnamefont {Kim}},\ }\href {\doibase 10.1103/PhysRevB.86.224417} {\bibfield
   {journal} {\bibinfo  {journal} {Phys. Rev. B}\ }\textbf {\bibinfo {volume}
  {86}},\ \bibinfo {pages} {224417} (\bibinfo {year} {2012})}\BibitemShut
  {NoStop}%
\bibitem [{\citenamefont {Price}\ and\ \citenamefont
  {Perkins}(2013)}]{price2013finite}%
  \BibitemOpen
  \bibfield  {author} {\bibinfo {author} {\bibfnamefont {C.}~\bibnamefont
  {Price}}\ and\ \bibinfo {author} {\bibfnamefont {N.~B.}\ \bibnamefont
  {Perkins}},\ }\href {\doibase 10.1103/PhysRevB.88.024410} {\bibfield
  {journal} {\bibinfo  {journal} {Phys. Rev. B}\ }\textbf {\bibinfo {volume}
  {88}},\ \bibinfo {pages} {024410} (\bibinfo {year} {2013})}\BibitemShut
  {NoStop}%
\bibitem [{\citenamefont {Rau}\ \emph {et~al.}(2014)\citenamefont {Rau},
  \citenamefont {Lee},\ and\ \citenamefont {Kee}}]{rau2014generic}%
  \BibitemOpen
  \bibfield  {author} {\bibinfo {author} {\bibfnamefont {J.~G.}\ \bibnamefont
  {Rau}}, \bibinfo {author} {\bibfnamefont {E.~K.-H.}\ \bibnamefont {Lee}}, \
  and\ \bibinfo {author} {\bibfnamefont {H.-Y.}\ \bibnamefont {Kee}},\ }\href
  {\doibase 10.1103/PhysRevLett.112.077204} {\bibfield  {journal} {\bibinfo
  {journal} {Phys. Rev. Lett.}\ }\textbf {\bibinfo {volume} {112}},\ \bibinfo
  {pages} {077204} (\bibinfo {year} {2014})}\BibitemShut {NoStop}%
\bibitem [{\citenamefont {Katukuri}\ \emph {et~al.}(2014)\citenamefont
  {Katukuri}, \citenamefont {Nishimoto}, \citenamefont {Yushankhai},
  \citenamefont {Stoyanova}, \citenamefont {Kandpal}, \citenamefont {Choi},
  \citenamefont {Coldea}, \citenamefont {Rousochatzakis}, \citenamefont
  {Hozoi},\ and\ \citenamefont {van~den Brink}}]{katukuri2014kitaev}%
  \BibitemOpen
  \bibfield  {author} {\bibinfo {author} {\bibfnamefont {V.~M.}\ \bibnamefont
  {Katukuri}}, \bibinfo {author} {\bibfnamefont {S.}~\bibnamefont {Nishimoto}},
  \bibinfo {author} {\bibfnamefont {V.}~\bibnamefont {Yushankhai}}, \bibinfo
  {author} {\bibfnamefont {A.}~\bibnamefont {Stoyanova}}, \bibinfo {author}
  {\bibfnamefont {H.}~\bibnamefont {Kandpal}}, \bibinfo {author} {\bibfnamefont
  {S.}~\bibnamefont {Choi}}, \bibinfo {author} {\bibfnamefont {R.}~\bibnamefont
  {Coldea}}, \bibinfo {author} {\bibfnamefont {I.}~\bibnamefont
  {Rousochatzakis}}, \bibinfo {author} {\bibfnamefont {L.}~\bibnamefont
  {Hozoi}}, \ and\ \bibinfo {author} {\bibfnamefont {J.}~\bibnamefont {van~den
  Brink}},\ }\href {http://iopscience.iop.org/1367-2630/16/1/013056} {\bibfield
   {journal} {\bibinfo  {journal} {New Journal of Physics}\ }\textbf {\bibinfo
  {volume} {16}},\ \bibinfo {pages} {013056} (\bibinfo {year}
  {2014})}\BibitemShut {NoStop}%
\bibitem [{\citenamefont {Choi}\ \emph {et~al.}(2012)\citenamefont {Choi},
  \citenamefont {Coldea}, \citenamefont {Kolmogorov}, \citenamefont
  {Lancaster}, \citenamefont {Mazin}, \citenamefont {Blundell}, \citenamefont
  {Radaelli}, \citenamefont {Singh}, \citenamefont {Gegenwart}, \citenamefont
  {Choi}, \citenamefont {Cheong}, \citenamefont {Baker}, \citenamefont
  {Stock},\ and\ \citenamefont {Taylor}}]{choi2012spin}%
  \BibitemOpen
  \bibfield  {author} {\bibinfo {author} {\bibfnamefont {S.~K.}\ \bibnamefont
  {Choi}}, \bibinfo {author} {\bibfnamefont {R.}~\bibnamefont {Coldea}},
  \bibinfo {author} {\bibfnamefont {A.~N.}\ \bibnamefont {Kolmogorov}},
  \bibinfo {author} {\bibfnamefont {T.}~\bibnamefont {Lancaster}}, \bibinfo
  {author} {\bibfnamefont {I.~I.}\ \bibnamefont {Mazin}}, \bibinfo {author}
  {\bibfnamefont {S.~J.}\ \bibnamefont {Blundell}}, \bibinfo {author}
  {\bibfnamefont {P.~G.}\ \bibnamefont {Radaelli}}, \bibinfo {author}
  {\bibfnamefont {Y.}~\bibnamefont {Singh}}, \bibinfo {author} {\bibfnamefont
  {P.}~\bibnamefont {Gegenwart}}, \bibinfo {author} {\bibfnamefont {K.~R.}\
  \bibnamefont {Choi}}, \bibinfo {author} {\bibfnamefont {S.-W.}\ \bibnamefont
  {Cheong}}, \bibinfo {author} {\bibfnamefont {P.~J.}\ \bibnamefont {Baker}},
  \bibinfo {author} {\bibfnamefont {C.}~\bibnamefont {Stock}}, \ and\ \bibinfo
  {author} {\bibfnamefont {J.}~\bibnamefont {Taylor}},\ }\href {\doibase
  10.1103/PhysRevLett.108.127204} {\bibfield  {journal} {\bibinfo  {journal}
  {Phys. Rev. Lett.}\ }\textbf {\bibinfo {volume} {108}},\ \bibinfo {pages}
  {127204} (\bibinfo {year} {2012})}\BibitemShut {NoStop}%
\bibitem [{\citenamefont {Ye}\ \emph {et~al.}(2012)\citenamefont {Ye},
  \citenamefont {Chi}, \citenamefont {Cao}, \citenamefont {Chakoumakos},
  \citenamefont {Fernandez-Baca}, \citenamefont {Custelcean}, \citenamefont
  {Qi}, \citenamefont {Korneta},\ and\ \citenamefont {Cao}}]{ye2012direct}%
  \BibitemOpen
  \bibfield  {author} {\bibinfo {author} {\bibfnamefont {F.}~\bibnamefont
  {Ye}}, \bibinfo {author} {\bibfnamefont {S.}~\bibnamefont {Chi}}, \bibinfo
  {author} {\bibfnamefont {H.}~\bibnamefont {Cao}}, \bibinfo {author}
  {\bibfnamefont {B.~C.}\ \bibnamefont {Chakoumakos}}, \bibinfo {author}
  {\bibfnamefont {J.~A.}\ \bibnamefont {Fernandez-Baca}}, \bibinfo {author}
  {\bibfnamefont {R.}~\bibnamefont {Custelcean}}, \bibinfo {author}
  {\bibfnamefont {T.~F.}\ \bibnamefont {Qi}}, \bibinfo {author} {\bibfnamefont
  {O.~B.}\ \bibnamefont {Korneta}}, \ and\ \bibinfo {author} {\bibfnamefont
  {G.}~\bibnamefont {Cao}},\ }\href {\doibase 10.1103/PhysRevB.85.180403}
  {\bibfield  {journal} {\bibinfo  {journal} {Phys. Rev. B}\ }\textbf {\bibinfo
  {volume} {85}},\ \bibinfo {pages} {180403} (\bibinfo {year}
  {2012})}\BibitemShut {NoStop}%
\bibitem [{\citenamefont {Comin}\ \emph {et~al.}(2012)\citenamefont {Comin},
  \citenamefont {Levy}, \citenamefont {Ludbrook}, \citenamefont {Zhu},
  \citenamefont {Veenstra}, \citenamefont {Rosen}, \citenamefont {Singh},
  \citenamefont {Gegenwart}, \citenamefont {Stricker}, \citenamefont {Hancock},
  \citenamefont {van~der Marel}, \citenamefont {Elfimov},\ and\ \citenamefont
  {Damascelli}}]{comin2012na}%
  \BibitemOpen
  \bibfield  {author} {\bibinfo {author} {\bibfnamefont {R.}~\bibnamefont
  {Comin}}, \bibinfo {author} {\bibfnamefont {G.}~\bibnamefont {Levy}},
  \bibinfo {author} {\bibfnamefont {B.}~\bibnamefont {Ludbrook}}, \bibinfo
  {author} {\bibfnamefont {Z.-H.}\ \bibnamefont {Zhu}}, \bibinfo {author}
  {\bibfnamefont {C.~N.}\ \bibnamefont {Veenstra}}, \bibinfo {author}
  {\bibfnamefont {J.~A.}\ \bibnamefont {Rosen}}, \bibinfo {author}
  {\bibfnamefont {Y.}~\bibnamefont {Singh}}, \bibinfo {author} {\bibfnamefont
  {P.}~\bibnamefont {Gegenwart}}, \bibinfo {author} {\bibfnamefont
  {D.}~\bibnamefont {Stricker}}, \bibinfo {author} {\bibfnamefont {J.~N.}\
  \bibnamefont {Hancock}}, \bibinfo {author} {\bibfnamefont {D.}~\bibnamefont
  {van~der Marel}}, \bibinfo {author} {\bibfnamefont {I.~S.}\ \bibnamefont
  {Elfimov}}, \ and\ \bibinfo {author} {\bibfnamefont {A.}~\bibnamefont
  {Damascelli}},\ }\href {\doibase 10.1103/PhysRevLett.109.266406} {\bibfield
  {journal} {\bibinfo  {journal} {Phys. Rev. Lett.}\ }\textbf {\bibinfo
  {volume} {109}},\ \bibinfo {pages} {266406} (\bibinfo {year}
  {2012})}\BibitemShut {NoStop}%
\bibitem [{\citenamefont {Gretarsson}\ \emph {et~al.}(2013)\citenamefont
  {Gretarsson}, \citenamefont {Clancy}, \citenamefont {Singh}, \citenamefont
  {Gegenwart}, \citenamefont {Hill}, \citenamefont {Kim}, \citenamefont
  {Upton}, \citenamefont {Said}, \citenamefont {Casa}, \citenamefont {Gog},\
  and\ \citenamefont {Kim}}]{gretarsson2013magnetic}%
  \BibitemOpen
  \bibfield  {author} {\bibinfo {author} {\bibfnamefont {H.}~\bibnamefont
  {Gretarsson}}, \bibinfo {author} {\bibfnamefont {J.~P.}\ \bibnamefont
  {Clancy}}, \bibinfo {author} {\bibfnamefont {Y.}~\bibnamefont {Singh}},
  \bibinfo {author} {\bibfnamefont {P.}~\bibnamefont {Gegenwart}}, \bibinfo
  {author} {\bibfnamefont {J.~P.}\ \bibnamefont {Hill}}, \bibinfo {author}
  {\bibfnamefont {J.}~\bibnamefont {Kim}}, \bibinfo {author} {\bibfnamefont
  {M.~H.}\ \bibnamefont {Upton}}, \bibinfo {author} {\bibfnamefont {A.~H.}\
  \bibnamefont {Said}}, \bibinfo {author} {\bibfnamefont {D.}~\bibnamefont
  {Casa}}, \bibinfo {author} {\bibfnamefont {T.}~\bibnamefont {Gog}}, \ and\
  \bibinfo {author} {\bibfnamefont {Y.-J.}\ \bibnamefont {Kim}},\ }\href
  {\doibase 10.1103/PhysRevB.87.220407} {\bibfield  {journal} {\bibinfo
  {journal} {Phys. Rev. B}\ }\textbf {\bibinfo {volume} {87}},\ \bibinfo
  {pages} {220407} (\bibinfo {year} {2013})}\BibitemShut {NoStop}%
\bibitem [{\citenamefont {Cao}\ \emph {et~al.}(2013)\citenamefont {Cao},
  \citenamefont {Qi}, \citenamefont {Li}, \citenamefont {Terzic}, \citenamefont
  {Cao}, \citenamefont {Yuan}, \citenamefont {Tovar}, \citenamefont {Murthy},\
  and\ \citenamefont {Kaul}}]{cao2013evolution}%
  \BibitemOpen
  \bibfield  {author} {\bibinfo {author} {\bibfnamefont {G.}~\bibnamefont
  {Cao}}, \bibinfo {author} {\bibfnamefont {T.~F.}\ \bibnamefont {Qi}},
  \bibinfo {author} {\bibfnamefont {L.}~\bibnamefont {Li}}, \bibinfo {author}
  {\bibfnamefont {J.}~\bibnamefont {Terzic}}, \bibinfo {author} {\bibfnamefont
  {V.~S.}\ \bibnamefont {Cao}}, \bibinfo {author} {\bibfnamefont {S.~J.}\
  \bibnamefont {Yuan}}, \bibinfo {author} {\bibfnamefont {M.}~\bibnamefont
  {Tovar}}, \bibinfo {author} {\bibfnamefont {G.}~\bibnamefont {Murthy}}, \
  and\ \bibinfo {author} {\bibfnamefont {R.~K.}\ \bibnamefont {Kaul}},\ }\href
  {\doibase 10.1103/PhysRevB.88.220414} {\bibfield  {journal} {\bibinfo
  {journal} {Phys. Rev. B}\ }\textbf {\bibinfo {volume} {88}},\ \bibinfo
  {pages} {220414} (\bibinfo {year} {2013})}\BibitemShut {NoStop}%
\bibitem [{\citenamefont {Manni}\ \emph {et~al.}(2014)\citenamefont {Manni},
  \citenamefont {Choi}, \citenamefont {Mazin}, \citenamefont {Coldea},
  \citenamefont {Altmeyer}, \citenamefont {Jeschke}, \citenamefont
  {Valent\'\i},\ and\ \citenamefont {Gegenwart}}]{manni2014effect}%
  \BibitemOpen
  \bibfield  {author} {\bibinfo {author} {\bibfnamefont {S.}~\bibnamefont
  {Manni}}, \bibinfo {author} {\bibfnamefont {S.}~\bibnamefont {Choi}},
  \bibinfo {author} {\bibfnamefont {I.~I.}\ \bibnamefont {Mazin}}, \bibinfo
  {author} {\bibfnamefont {R.}~\bibnamefont {Coldea}}, \bibinfo {author}
  {\bibfnamefont {M.}~\bibnamefont {Altmeyer}}, \bibinfo {author}
  {\bibfnamefont {H.~O.}\ \bibnamefont {Jeschke}}, \bibinfo {author}
  {\bibfnamefont {R.}~\bibnamefont {Valent\'\i}}, \ and\ \bibinfo {author}
  {\bibfnamefont {P.}~\bibnamefont {Gegenwart}},\ }\href {\doibase
  10.1103/PhysRevB.89.245113} {\bibfield  {journal} {\bibinfo  {journal} {Phys.
  Rev. B}\ }\textbf {\bibinfo {volume} {89}},\ \bibinfo {pages} {245113}
  (\bibinfo {year} {2014})}\BibitemShut {NoStop}%
\bibitem [{\citenamefont {Knolle}\ \emph {et~al.}(2014)\citenamefont {Knolle},
  \citenamefont {Chern}, \citenamefont {Kovrizhin}, \citenamefont {Moessner},\
  and\ \citenamefont {Perkins}}]{knolle2014raman}%
  \BibitemOpen
  \bibfield  {author} {\bibinfo {author} {\bibfnamefont {J.}~\bibnamefont
  {Knolle}}, \bibinfo {author} {\bibfnamefont {G.-W.}\ \bibnamefont {Chern}},
  \bibinfo {author} {\bibfnamefont {D.}~\bibnamefont {Kovrizhin}}, \bibinfo
  {author} {\bibfnamefont {R.}~\bibnamefont {Moessner}}, \ and\ \bibinfo
  {author} {\bibfnamefont {N.}~\bibnamefont {Perkins}},\ }\href@noop {}
  {\bibfield  {journal} {\bibinfo  {journal} {Physical review letters}\
  }\textbf {\bibinfo {volume} {113}},\ \bibinfo {pages} {187201} (\bibinfo
  {year} {2014})}\BibitemShut {NoStop}%
\bibitem [{\citenamefont {Chun}\ \emph {et~al.}(2015)\citenamefont {Chun},
  \citenamefont {Kim}, \citenamefont {Kim}, \citenamefont {Zheng},
  \citenamefont {Stoumpos}, \citenamefont {Malliakas}, \citenamefont
  {Mitchell}, \citenamefont {Mehlawat}, \citenamefont {Singh}, \citenamefont
  {Choi} \emph {et~al.}}]{chun2015direct}%
  \BibitemOpen
  \bibfield  {author} {\bibinfo {author} {\bibfnamefont {S.~H.}\ \bibnamefont
  {Chun}}, \bibinfo {author} {\bibfnamefont {J.-W.}\ \bibnamefont {Kim}},
  \bibinfo {author} {\bibfnamefont {J.}~\bibnamefont {Kim}}, \bibinfo {author}
  {\bibfnamefont {H.}~\bibnamefont {Zheng}}, \bibinfo {author} {\bibfnamefont
  {C.~C.}\ \bibnamefont {Stoumpos}}, \bibinfo {author} {\bibfnamefont
  {C.}~\bibnamefont {Malliakas}}, \bibinfo {author} {\bibfnamefont
  {J.}~\bibnamefont {Mitchell}}, \bibinfo {author} {\bibfnamefont
  {K.}~\bibnamefont {Mehlawat}}, \bibinfo {author} {\bibfnamefont
  {Y.}~\bibnamefont {Singh}}, \bibinfo {author} {\bibfnamefont
  {Y.}~\bibnamefont {Choi}},  \emph {et~al.},\ }\href@noop {} {\bibfield
  {journal} {\bibinfo  {journal} {Nature Physics}\ } (\bibinfo {year}
  {2015})}\BibitemShut {NoStop}%
\bibitem [{\citenamefont {Takayama}\ \emph {et~al.}(2015)\citenamefont
  {Takayama}, \citenamefont {Kato}, \citenamefont {Dinnebier}, \citenamefont
  {Nuss}, \citenamefont {Kono}, \citenamefont {Veiga}, \citenamefont {Fabbris},
  \citenamefont {Haskel},\ and\ \citenamefont
  {Takagi}}]{takayama2015hyperhoneycomb}%
  \BibitemOpen
  \bibfield  {author} {\bibinfo {author} {\bibfnamefont {T.}~\bibnamefont
  {Takayama}}, \bibinfo {author} {\bibfnamefont {A.}~\bibnamefont {Kato}},
  \bibinfo {author} {\bibfnamefont {R.}~\bibnamefont {Dinnebier}}, \bibinfo
  {author} {\bibfnamefont {J.}~\bibnamefont {Nuss}}, \bibinfo {author}
  {\bibfnamefont {H.}~\bibnamefont {Kono}}, \bibinfo {author} {\bibfnamefont
  {L.~S.~I.}\ \bibnamefont {Veiga}}, \bibinfo {author} {\bibfnamefont
  {G.}~\bibnamefont {Fabbris}}, \bibinfo {author} {\bibfnamefont
  {D.}~\bibnamefont {Haskel}}, \ and\ \bibinfo {author} {\bibfnamefont
  {H.}~\bibnamefont {Takagi}},\ }\href {\doibase
  10.1103/PhysRevLett.114.077202} {\bibfield  {journal} {\bibinfo  {journal}
  {Phys. Rev. Lett.}\ }\textbf {\bibinfo {volume} {114}},\ \bibinfo {pages}
  {077202} (\bibinfo {year} {2015})}\BibitemShut {NoStop}%
\bibitem [{\citenamefont {Modic}\ \emph {et~al.}(2014)\citenamefont {Modic},
  \citenamefont {Smidt}, \citenamefont {Kimchi}, \citenamefont {Breznay},
  \citenamefont {Biffin}, \citenamefont {Choi}, \citenamefont {Johnson},
  \citenamefont {Coldea}, \citenamefont {Watkins-Curry}, \citenamefont
  {McCandless} \emph {et~al.}}]{modic2014realization}%
  \BibitemOpen
  \bibfield  {author} {\bibinfo {author} {\bibfnamefont {K.}~\bibnamefont
  {Modic}}, \bibinfo {author} {\bibfnamefont {T.~E.}\ \bibnamefont {Smidt}},
  \bibinfo {author} {\bibfnamefont {I.}~\bibnamefont {Kimchi}}, \bibinfo
  {author} {\bibfnamefont {N.~P.}\ \bibnamefont {Breznay}}, \bibinfo {author}
  {\bibfnamefont {A.}~\bibnamefont {Biffin}}, \bibinfo {author} {\bibfnamefont
  {S.}~\bibnamefont {Choi}}, \bibinfo {author} {\bibfnamefont {R.~D.}\
  \bibnamefont {Johnson}}, \bibinfo {author} {\bibfnamefont {R.}~\bibnamefont
  {Coldea}}, \bibinfo {author} {\bibfnamefont {P.}~\bibnamefont
  {Watkins-Curry}}, \bibinfo {author} {\bibfnamefont {G.~T.}\ \bibnamefont
  {McCandless}},  \emph {et~al.},\ }\href
  {http://www.nature.com/ncomms/2014/140627/ncomms5203/full/ncomms5203.html}
  {\bibfield  {journal} {\bibinfo  {journal} {Nature communications}\ }\textbf
  {\bibinfo {volume} {5}} (\bibinfo {year} {2014})}\BibitemShut {NoStop}%
\bibitem [{\citenamefont {Mandal}\ and\ \citenamefont
  {Surendran}(2009)}]{mandal2009exactly}%
  \BibitemOpen
  \bibfield  {author} {\bibinfo {author} {\bibfnamefont {S.}~\bibnamefont
  {Mandal}}\ and\ \bibinfo {author} {\bibfnamefont {N.}~\bibnamefont
  {Surendran}},\ }\href {\doibase 10.1103/PhysRevB.79.024426} {\bibfield
  {journal} {\bibinfo  {journal} {Phys. Rev. B}\ }\textbf {\bibinfo {volume}
  {79}},\ \bibinfo {pages} {024426} (\bibinfo {year} {2009})}\BibitemShut
  {NoStop}%
\bibitem [{\citenamefont {Lee}\ \emph {et~al.}(2014{\natexlab{a}})\citenamefont
  {Lee}, \citenamefont {Schaffer}, \citenamefont {Bhattacharjee},\ and\
  \citenamefont {Kim}}]{lee2014heisenberg}%
  \BibitemOpen
  \bibfield  {author} {\bibinfo {author} {\bibfnamefont {E.~K.-H.}\
  \bibnamefont {Lee}}, \bibinfo {author} {\bibfnamefont {R.}~\bibnamefont
  {Schaffer}}, \bibinfo {author} {\bibfnamefont {S.}~\bibnamefont
  {Bhattacharjee}}, \ and\ \bibinfo {author} {\bibfnamefont {Y.~B.}\
  \bibnamefont {Kim}},\ }\href {\doibase 10.1103/PhysRevB.89.045117} {\bibfield
   {journal} {\bibinfo  {journal} {Phys. Rev. B}\ }\textbf {\bibinfo {volume}
  {89}},\ \bibinfo {pages} {045117} (\bibinfo {year}
  {2014}{\natexlab{a}})}\BibitemShut {NoStop}%
\bibitem [{\citenamefont {Lee}\ \emph {et~al.}(2014{\natexlab{b}})\citenamefont
  {Lee}, \citenamefont {Lee}, \citenamefont {Paramekanti},\ and\ \citenamefont
  {Kim}}]{lee2014order}%
  \BibitemOpen
  \bibfield  {author} {\bibinfo {author} {\bibfnamefont {S.}~\bibnamefont
  {Lee}}, \bibinfo {author} {\bibfnamefont {E.~K.-H.}\ \bibnamefont {Lee}},
  \bibinfo {author} {\bibfnamefont {A.}~\bibnamefont {Paramekanti}}, \ and\
  \bibinfo {author} {\bibfnamefont {Y.~B.}\ \bibnamefont {Kim}},\ }\href
  {\doibase 10.1103/PhysRevB.89.014424} {\bibfield  {journal} {\bibinfo
  {journal} {Phys. Rev. B}\ }\textbf {\bibinfo {volume} {89}},\ \bibinfo
  {pages} {014424} (\bibinfo {year} {2014}{\natexlab{b}})}\BibitemShut
  {NoStop}%
\bibitem [{\citenamefont {Lee}\ \emph {et~al.}(2014{\natexlab{c}})\citenamefont
  {Lee}, \citenamefont {Bhattacharjee}, \citenamefont {Hwang}, \citenamefont
  {Kim}, \citenamefont {Jin},\ and\ \citenamefont {Kim}}]{lee2014topological}%
  \BibitemOpen
  \bibfield  {author} {\bibinfo {author} {\bibfnamefont {E.~K.-H.}\
  \bibnamefont {Lee}}, \bibinfo {author} {\bibfnamefont {S.}~\bibnamefont
  {Bhattacharjee}}, \bibinfo {author} {\bibfnamefont {K.}~\bibnamefont
  {Hwang}}, \bibinfo {author} {\bibfnamefont {H.-S.}\ \bibnamefont {Kim}},
  \bibinfo {author} {\bibfnamefont {H.}~\bibnamefont {Jin}}, \ and\ \bibinfo
  {author} {\bibfnamefont {Y.~B.}\ \bibnamefont {Kim}},\ }\href {\doibase
  10.1103/PhysRevB.89.205132} {\bibfield  {journal} {\bibinfo  {journal} {Phys.
  Rev. B}\ }\textbf {\bibinfo {volume} {89}},\ \bibinfo {pages} {205132}
  (\bibinfo {year} {2014}{\natexlab{c}})}\BibitemShut {NoStop}%
\bibitem [{\citenamefont {Nasu}\ \emph {et~al.}(2014)\citenamefont {Nasu},
  \citenamefont {Udagawa},\ and\ \citenamefont
  {Motome}}]{nasu2014vaporization}%
  \BibitemOpen
  \bibfield  {author} {\bibinfo {author} {\bibfnamefont {J.}~\bibnamefont
  {Nasu}}, \bibinfo {author} {\bibfnamefont {M.}~\bibnamefont {Udagawa}}, \
  and\ \bibinfo {author} {\bibfnamefont {Y.}~\bibnamefont {Motome}},\ }\href
  {http://arxiv.org/abs/1406.5415} {\bibfield  {journal} {\bibinfo  {journal}
  {arXiv preprint arXiv:1406.5415}\ } (\bibinfo {year} {2014})}\BibitemShut
  {NoStop}%
\bibitem [{\citenamefont {Lee}\ and\ \citenamefont
  {Kim}(2015)}]{lee2015theory}%
  \BibitemOpen
  \bibfield  {author} {\bibinfo {author} {\bibfnamefont {E.~K.-H.}\
  \bibnamefont {Lee}}\ and\ \bibinfo {author} {\bibfnamefont {Y.~B.}\
  \bibnamefont {Kim}},\ }\href {\doibase 10.1103/PhysRevB.91.064407} {\bibfield
   {journal} {\bibinfo  {journal} {Phys. Rev. B}\ }\textbf {\bibinfo {volume}
  {91}},\ \bibinfo {pages} {064407} (\bibinfo {year} {2015})}\BibitemShut
  {NoStop}%
\bibitem [{\citenamefont {Kimchi}\ \emph {et~al.}(2014)\citenamefont {Kimchi},
  \citenamefont {Coldea},\ and\ \citenamefont
  {Vishwanath}}]{kimchi2014unified}%
  \BibitemOpen
  \bibfield  {author} {\bibinfo {author} {\bibfnamefont {I.}~\bibnamefont
  {Kimchi}}, \bibinfo {author} {\bibfnamefont {R.}~\bibnamefont {Coldea}}, \
  and\ \bibinfo {author} {\bibfnamefont {A.}~\bibnamefont {Vishwanath}},\
  }\href {http://arxiv.org/abs/1408.3640} {\bibfield  {journal} {\bibinfo
  {journal} {arXiv preprint arXiv:1408.3640}\ } (\bibinfo {year}
  {2014})}\BibitemShut {NoStop}%
\bibitem [{\citenamefont {Kim}\ \emph {et~al.}(2015)\citenamefont {Kim},
  \citenamefont {Lee},\ and\ \citenamefont {Kim}}]{kim2015predominance}%
  \BibitemOpen
  \bibfield  {author} {\bibinfo {author} {\bibfnamefont {H.-S.}\ \bibnamefont
  {Kim}}, \bibinfo {author} {\bibfnamefont {E.~K.-H.}\ \bibnamefont {Lee}}, \
  and\ \bibinfo {author} {\bibfnamefont {Y.~B.}\ \bibnamefont {Kim}},\
  }\href@noop {} {\bibfield  {journal} {\bibinfo  {journal} {arXiv preprint
  arXiv:1502.00006}\ } (\bibinfo {year} {2015})}\BibitemShut {NoStop}%
\bibitem [{\citenamefont {Schaffer}\ \emph {et~al.}(2015)\citenamefont
  {Schaffer}, \citenamefont {Lee}, \citenamefont {Lu},\ and\ \citenamefont
  {Kim}}]{schaffer2015topological}%
  \BibitemOpen
  \bibfield  {author} {\bibinfo {author} {\bibfnamefont {R.}~\bibnamefont
  {Schaffer}}, \bibinfo {author} {\bibfnamefont {E.~K.-H.}\ \bibnamefont
  {Lee}}, \bibinfo {author} {\bibfnamefont {Y.-M.}\ \bibnamefont {Lu}}, \ and\
  \bibinfo {author} {\bibfnamefont {Y.~B.}\ \bibnamefont {Kim}},\ }\href
  {\doibase 10.1103/PhysRevLett.114.116803} {\bibfield  {journal} {\bibinfo
  {journal} {Phys. Rev. Lett.}\ }\textbf {\bibinfo {volume} {114}},\ \bibinfo
  {pages} {116803} (\bibinfo {year} {2015})}\BibitemShut {NoStop}%
\bibitem [{\citenamefont {Biffin}\ \emph
  {et~al.}(2014{\natexlab{a}})\citenamefont {Biffin}, \citenamefont {Johnson},
  \citenamefont {Choi}, \citenamefont {Freund}, \citenamefont {Manni},
  \citenamefont {Bombardi}, \citenamefont {Manuel}, \citenamefont {Gegenwart},\
  and\ \citenamefont {Coldea}}]{biffin2014unconventional}%
  \BibitemOpen
  \bibfield  {author} {\bibinfo {author} {\bibfnamefont {A.}~\bibnamefont
  {Biffin}}, \bibinfo {author} {\bibfnamefont {R.}~\bibnamefont {Johnson}},
  \bibinfo {author} {\bibfnamefont {S.}~\bibnamefont {Choi}}, \bibinfo {author}
  {\bibfnamefont {F.}~\bibnamefont {Freund}}, \bibinfo {author} {\bibfnamefont
  {S.}~\bibnamefont {Manni}}, \bibinfo {author} {\bibfnamefont
  {A.}~\bibnamefont {Bombardi}}, \bibinfo {author} {\bibfnamefont
  {P.}~\bibnamefont {Manuel}}, \bibinfo {author} {\bibfnamefont
  {P.}~\bibnamefont {Gegenwart}}, \ and\ \bibinfo {author} {\bibfnamefont
  {R.}~\bibnamefont {Coldea}},\ }\href@noop {} {\bibfield  {journal} {\bibinfo
  {journal} {Physical Review B}\ }\textbf {\bibinfo {volume} {90}},\ \bibinfo
  {pages} {205116} (\bibinfo {year} {2014}{\natexlab{a}})}\BibitemShut
  {NoStop}%
\bibitem [{\citenamefont {Biffin}\ \emph
  {et~al.}(2014{\natexlab{b}})\citenamefont {Biffin}, \citenamefont {Johnson},
  \citenamefont {Kimchi}, \citenamefont {Morris}, \citenamefont {Bombardi},
  \citenamefont {Analytis}, \citenamefont {Vishwanath},\ and\ \citenamefont
  {Coldea}}]{biffin2014noncoplanar}%
  \BibitemOpen
  \bibfield  {author} {\bibinfo {author} {\bibfnamefont {A.}~\bibnamefont
  {Biffin}}, \bibinfo {author} {\bibfnamefont {R.}~\bibnamefont {Johnson}},
  \bibinfo {author} {\bibfnamefont {I.}~\bibnamefont {Kimchi}}, \bibinfo
  {author} {\bibfnamefont {R.}~\bibnamefont {Morris}}, \bibinfo {author}
  {\bibfnamefont {A.}~\bibnamefont {Bombardi}}, \bibinfo {author}
  {\bibfnamefont {J.}~\bibnamefont {Analytis}}, \bibinfo {author}
  {\bibfnamefont {A.}~\bibnamefont {Vishwanath}}, \ and\ \bibinfo {author}
  {\bibfnamefont {R.}~\bibnamefont {Coldea}},\ }\href@noop {} {\bibfield
  {journal} {\bibinfo  {journal} {Physical review letters}\ }\textbf {\bibinfo
  {volume} {113}},\ \bibinfo {pages} {197201} (\bibinfo {year}
  {2014}{\natexlab{b}})}\BibitemShut {NoStop}%
\bibitem [{Note1()}]{Note1}%
  \BibitemOpen
  \bibinfo {note} {For further details on the crystal structure of these two 3D
  honeycomb iridates, we refer to earlier works that have elaborated on the
  description of the structure and crystal symmetries.\cite
  {modic2014realization, takayama2015hyperhoneycomb,
  lee2015theory}}\BibitemShut {NoStop}%
\bibitem [{ITe()}]{ITensor}%
  \BibitemOpen
  \href@noop {} {}\bibinfo {note} {Calculations were performed using the
  ITensor Library, http://itensor.org/}\BibitemShut {NoStop}%
\end{thebibliography}%

\newpage

\appendix

\section{\label{app:irrep}Magnetic basis vectors}
The irreducible representations and magenetic basis vectors for a
magnetic structure with ordering wavevector $\vec{q}=(h00)$ in the
orthorhombic unit cell of the hyper-honeycomb lattice are given
by\cite{biffin2014unconventional}
\begin{table}[h!]
\begin{tabular}{c|c}
  Irreducible & Basis\\
  Representation & Vectors \\
  \hline 
  $\Gamma_1$ & $F_a$, $G_b$, $A_c$ \\
  $\Gamma_2$ & $C_a$, $A_b$, $G_c$ \\
  $\Gamma_3$ & $G_a$, $F_b$, $C_c$ \\
  $\Gamma_4$ & $A_a$, $C_b$, $F_c$
\end{tabular}
\end{table}

In the sublattice basis given in Fig. \ref{fig:h0_lattice}, the
magnetic basis vectors are given by\cite{biffin2014unconventional}
\begin{equation}
  \label{eq:magbasisvecs}
  F=
  \begin{bmatrix}
    1 \\
    1 \\
    1 \\
    1 \\
  \end{bmatrix},
  C=
  \begin{bmatrix}
    1 \\
    1 \\
    -1 \\
    -1 \\
  \end{bmatrix},
  A=
  \begin{bmatrix}
    1 \\
    -1 \\
    -1 \\
    1 \\
  \end{bmatrix},
  G=
  \begin{bmatrix}
    1 \\
    -1 \\
    1 \\
    -1 \\
  \end{bmatrix}.
\end{equation}

With a structure specified by the basis vectors $\vec{v}=(i v_a,i
v_b,v_c)$, the moments at each site is given
by\cite{biffin2014unconventional}
\begin{align}
  \label{eq:mag}
  \vec{S}_n(\vec{r})&=\hat{a} S_a v_a(n) \sin \vec{q}\cdot \vec{r} + \hat{b} S_b v_b(n) \sin \vec{q}\cdot \vec{r} \nonumber \\
  &+\hat{c} S_c v_c(n) \cos \vec{q}\cdot \vec{r},
\end{align}
where $\hat{a}$, $\hat{b}$, $\hat{c}$ are the unit vectors of the
orthorhombic lattice vectors, $\vec{S}_n(\vec{r})$ is the moment at
position $\vec{r}$ which has sublattice index $n$, and $S_i$ are the
magnitudes of each component of the moment.  Since all phases
predicted in this work have relative phases between components given
by $\vec{v}=(i v_a,i v_b,v_c)$, we adopt a shorthand notation by
dropping the relative imaginary unit $i$:
$\vec{v}\equiv(v_a,v_b,v_c)$.

For the stripy-honeycomb, there are 8 sublattices in the unit cell as
opposed to 4 for the hyper-honeycomb.  We break up those 8
sublattices into two sets of 4---the unprimed $1-4$ sublattices and
the primed sublattices $1'-4'$ as seen in
Fig. \ref{fig:h1_lattice}.\cite{biffin2014noncoplanar} The irreducible
representations and magnetic basis vectors for wavevector $h00$ are
given by\cite{biffin2014noncoplanar}
\begin{table}[h!]
\begin{tabular}{c|c}
  Irreducible & Basis\\
  Representation & Vectors \\
  \hline 
  $\Gamma_1$ & $C_a$, $A_b$, $G_c$ \\
  $\Gamma_2$ & $F_a$, $G_b$, $A_c$ \\
  $\Gamma_3$ & $A_a$, $C_b$, $F_c$ \\
  $\Gamma_4$ & $G_a$, $F_b$, $C_c$
\end{tabular}
\end{table}

The magnetic basis vectors are also specified by
Eq. \ref{eq:magbasisvecs}.  The structure given by
\begin{equation}
  \label{eq:h1_basis}
  \vec{v}=i(v_a,-v'_a),i(v_b,-v'_b),(v_c,v'_c)  
\end{equation}
is also specified by Eq. \ref{eq:mag} but with a relative sign on the
$a$ and $b$ components between the primed and unprimed sublattices.
Since all magnetic structures predicted in this work has the relative
phases given by Eq. \ref{eq:h1_basis}, we adopt the shorthand notation
where $\vec{v}\equiv(v_a,v_b,v_c)$.

\section{\label{app:1d}1D spin chain results}

\begin{figure}[tp]
  \centering
  \setlength\fboxsep{0pt}
  \setlength\fboxrule{0pt}
  \subfloat[][$J_1 = 0.33$]{
    \fbox{\includegraphics[width=1.58in]{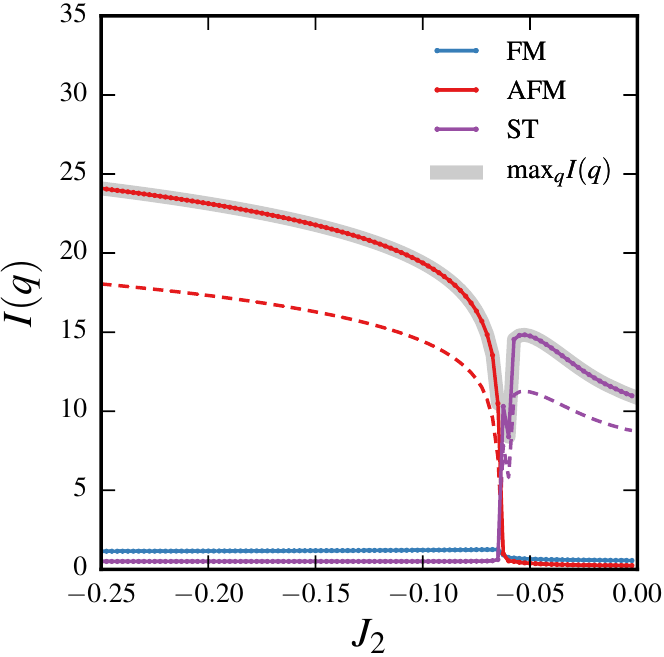}
    }
  }
  \subfloat[][$J_1 = 0.25$]{
    \fbox{\includegraphics[width=1.58in]{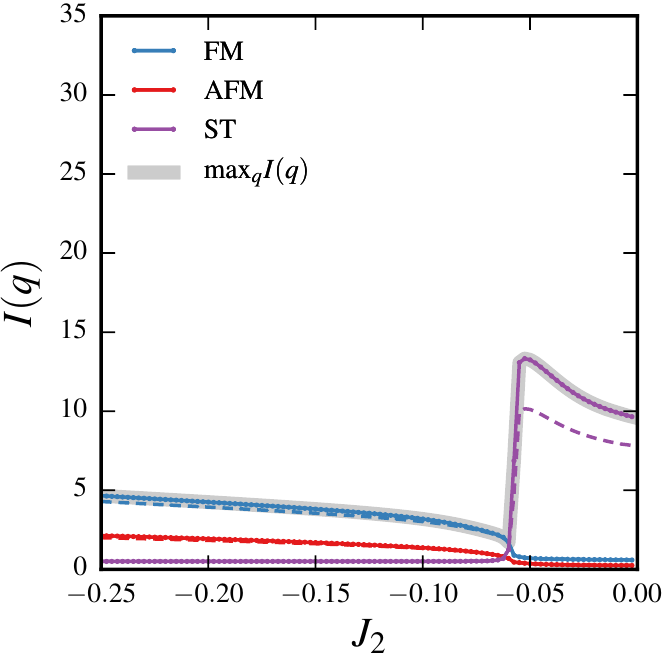}
    }
  }
  \caption{\label{fig:dmrg_corr} Decoupled CZC model: specific cuts of
    the static structure factor $I(q) = \sum_{\mu} I^{\mu\mu}(q)$ for
    $q=0$ (FM), $\pi/2$ (ST) and $\pi$ (AF) are shown in (a-b) for
    $N=128$ (solid lines) and $N=96$ (dashed lines). Along with these
    we shown the maximum of $I(q)$, which always belongs to one of
    these three choices.}
\end{figure}

\begin{figure}[tp]
  \centering
  \setlength\fboxsep{0pt}
  \setlength\fboxrule{0pt}
  \fbox{\includegraphics[width=2in]{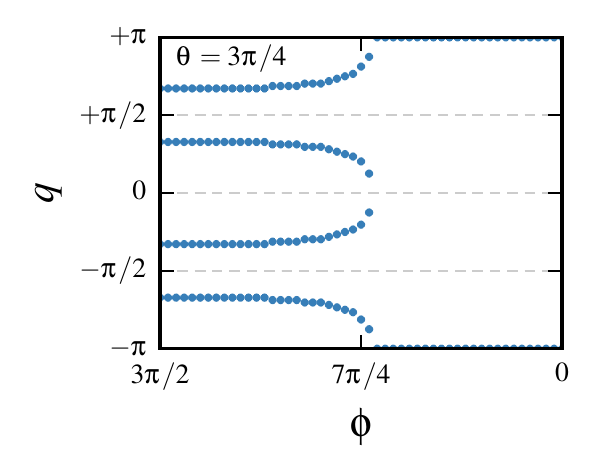}
  }
  \caption{\label{fig:1d_jkg_cut} Decoupled \jkg{} model: location of
    the maximum static structure factor (\textit{i.e.}
    $\text{max}_q[I(q) = \sum_{\mu} I^{\mu\mu}(q)]$) for
    $\theta=3\pi/4$ as a function of $\phi$ for the $N=128$ case.  The
    antiferromagnetic phase for $\phi \gtrsim 7\pi/4$ has the maximum
    peak located at $\pm \pi$.  Below this value of $\phi$,
    incommensurate correlations develop and persist down to $\phi =
    3\pi/2$.}
\end{figure}

To determine the ground state phase diagram of Eq. (\ref{eq:chain}),
we used the DMRG as implemented in the ITensor
framework.\cite{ITensor} We consider a chain of $N \leq 128$ sites
with open boundary conditions.  In the decoupled \jkg{} limit where
$J_2=0$, we parameterize our phase diagram in the same manner as per
Eq. \ref{eq:parametrization}.  In the decoupled CZC limit where
$\Gamma=0$, we fix the energy scale $K=-1$ such that there is a
two-dimensional parameter space $(J_1,J_2)$ to explore. For each point
in parameter space we perform 10-20 sweeps, keeping up to $1000$
states resulting in a maximum truncation error of $\sim 10^{-9}$. For
small system sizes ($N \leq 22$) these results were checked for
consistency against exact diagonalization.

To characterize the phases we consider two diagnostics: derivatives of
the ground state energy and the static structure factor.  Sharp
features in the energy derivatives signal phase boundaries while peaks
in the diagonal components of the static structure factor
\begin{equation}
  I^{\mu\mu}({q}) \equiv \frac{1}{N} \sum_{ij} e^{i {q} \cdot(i -j)} 
\avg{S^{\mu}_i S^{\mu}_j},
\end{equation}
signal the ordering wave-vector.

For the decoupled CZC model, we find that throughout the entire phase
diagram the maxima of the structure factor $I(q)$ appear only at
wave-vectors $q=0$, $\pi/2$ or $\pi$ corresponding to a ferromagnet
(FM-$xy$), antiferromagnet (AF-$z$) or stripy (ST) phase. The
isotropic structure factor $I \equiv \sum_{\mu} I^{\mu\mu}(q)$ for
each of these wave-vectors is shown in Fig. \ref{fig:dmrg_corr} for
both $N=96$ and $N=128$. The ferromagnet and stripy phases primarily
have correlations in the $\hat{x}$ and $\hat{y}$ directions with the
maxima in appearing in $I^{xx}$ or $I^{yy}$, while the
anti-ferromagnet is oriented in the $\hat{z}$ direction, with the
corresponding maximum in $I^{zz}$.

The gross features of this phase diagram can be understood easily.
First we have two exactly solvable points present in the $J_2=0$
limit; the Kitaev chain \cite{} at $J_1=0$ and a point dual to a
Heisenberg ferromagnet \cite{} at $J_1=1/2$.  As in the
two-dimensional case, a stripy phase extends between these two
solvable points. This is stable to finite $J_2$, but ultimately gives
way to either the FM-$xy$ or the AF-$z$ phases. These can be
understood by considering the large $J_2$ limit where the even and odd
sites decouple into two ferromagnetic chains. The relative orientation
of the two sublattices is then determined by the $K$ and $J_1$
interactions, with $K < 0$ favouring ferromagnetism and $J_1 > 0$
favouring anti-ferromagnetism.

The anti-ferromagnetic phase for $J_1 \gtrsim 0.3$ can also be
understood by considering the solvable point at $(J_1,J_2) = (1/2,0)$.
This point is dual to the ferromagnetic Heisenberg point via a
four-sublattice transformation, and hence possesses SU$(2)$ spin
rotation symmetry in the rotated basis.  One finds this continuous
degeneracy is broken by finite $J_2$, selecting the AF-$z$ state as
the ground state, which is a simple product state.  Since this product
state is the ground state of the ferromagnetic $J_2$ interaction and
the $(J_1,K)=(1/2,-1)$ interaction independently, it follows that this
product state is the exact ground state of the
$(J_1,K,J_2)=(1/2,-1,J_2)$ line.  As $J_1$ decreases, the AF
correlations are preserved but the ground state is no longer a simple
product state.

For the decoupled \jkg{} model, we find one boundary across the phase
diagram as indicated by energy derivatives.  The maxima of the
structure factor of the phase connected to the antiferromagnetic
Heisenberg point is located at $\pm \pi$, indicating AF correlations.
As we move across the boundary to the adjacent region, peaks in the
structure factor develop at incommensurate wavevectors while the peak
at $\pm \pi$ vanishes.  This incommensurate peak's position varies as
we approach the $J_1=0$ limit.  The stripy phase located in the
$\Gamma=0$ limit does not extend into the finite $\Gamma$ region, in
contrast to the effects of $J_2$ in the decoupled CZC model.  We
illustrate the position of the maximum of the static structure factor
for $\theta=3\pi/4$ as a function of $\phi$ in
Fig. \ref{fig:1d_jkg_cut}.
\end{document}